\documentclass[a4paper,11pt]{article}
\pdfoutput=1 % if your are submitting a pdflatex (i.e. if you have
\usepackage{jcappub} % for details on the use of the package, please
\usepackage[T1]{fontenc} % if needed
\usepackage{epsfig}

\usepackage{amssymb}
\usepackage{amsthm}
\usepackage{amsmath}
\usepackage{amsbsy}
\usepackage{graphicx}
\usepackage{caption}
\usepackage{subcaption}
\usepackage{float}
\usepackage{lineno}

\title{\boldmath A lower bound on the number of cosmic ray events required to measure source catalogue correlations.}

\author[a,b]{Marco Dolci}
\author[a]{Andrew Romero-Wolf}
\author[c,d]{Stephanie Wissel}

\affiliation[a]{Jet Propulsion Laboratory, California Institute of Technology, Pasadena, California 91109, USA}
\affiliation[b]{Department of Mechanical and Aerospace Engineering, Politecnico di Torino, Turin, Italy}
\affiliation[c]{University of California, Los Angeles, California 90095, USA}
\affiliation[d]{California Polytechnic State University, San Luis Obispo, California 93407, USA}

\emailAdd{marco.dolci@polito.it}
\emailAdd{Andrew.Romero-Wolf@jpl.nasa.gov}
\emailAdd{swissel@calpoly.edu}

\abstract{Recent analyses of cosmic ray arrival directions have resulted in evidence for a positive correlation with active galactic nuclei positions that has weak significance against an isotropic source distribution. In this paper, we explore the sample size needed to measure a highly statistically significant correlation to a parent source catalogue. We compare several scenarios for the directional scattering of ultra-high energy cosmic rays given our current knowledge of the galactic and intergalactic magnetic fields. We find significant correlations are possible for a sample of $>$1000 cosmic ray protons with energies above 60 EeV.}

\begin{document}
\maketitle
\flushbottom

\section{Introduction}
%\begin{itemize}
%\item Identifying UHECRs is hard because of a list of physical effects. % ARW ok.
%\item State of the art measurements. % ARW ok
%\item Future detectors: all-sky, increased exposure. % SAW ok
%\item What this paper aims to do.
%    \begin{itemize}
%    \item Sample Size needed to detect source catalogue correlation.
%    \item Do this as simply as possible: parametric analysis based on results of more detailed simulations.
%    \end{itemize}
%\item Previous literature along the lines of what we do.
%\item Organization of the paper.
%\end{itemize}
The origin of ultra-high energy cosmic rays (UHECRs) is a long-standing unsolved problem, which has defied an observational solution in large part due to magnetic field scattering. As cosmic rays propagate towards Earth, the galactic and intergalactic magnetic fields deflect their trajectories so that only the highest energy particles are rigid enough to point back to their sources. UHECRs also lose energy as they propagate through space due to interactions with photon backgrounds~\cite{greisen_1966, zatsepin_and_kuzmin_1966}, meaning that the sources of UHECRs observed with energies $\gtrsim 6 \times 10^{19}$ eV are expected to be nearby. The intergalactic magnetic fields, although weak, interact with UHECRs over the whole course of their trajectory, which can result in large deflections. These combined effects mean that arrival directions can become isotropized resulting in weakly significant correlations with a parent source catalogue.

Recently, the Auger collaboration has been searching for correlation in the arrival directions of ultra-high energy cosmic rays with the active galactic nuclei (AGN) distribution represented by the Veron-Cetty \& Veron (VCV) catalogue \cite{abraham_2007, abraham_2008, aab_2015, abreu_2010}. They compare their catalogue correlation with the correlation due to an isotropic source distribution to estimate the statistical significance of their result. The test assumes three parameters: 1) the energy threshold, motivated by the expectation that above the Greisen-Zatsepin-Kuzmin (GZK) cutoff~\cite{greisen_1966, zatsepin_and_kuzmin_1966} UHECRs are extra-galactic, 2) a correlation angle, motivated by the degree to which the galactic magnetic field may scatter events and result in significant source confusion, and 3) the maximum redshift of catalogue sources to consider in the correlation analysis, motivated by the expectation that UHECRs above the GZK cutoff are within a few hundred megaparsecs. 

In their latest release \cite{aab_2015}, the correlation to VCV AGN provided only weak evidence of anisotropy along with weak clustering around Centaurus A. The HiRes and Telescope Array collaborations performed similar studies finding no highly statistically significant correlation with a source catalogue~\cite{abbasi_2008, abuzayyad_2012, abuzayyad_2013}. 
%This sentence has little bearing on our study. It is a clustering analysis. SAW 
%However, a recent study of the clustering of events with the telescope array (TA) has found weak evidence of UHECR clustering ``hot-spot" at the $20^{\circ}$ scale~\cite{abbasi_2014}.

Future UHECR detectors could provide the exposure needed to reveal anisotropy and clustering to source catalogues. Proposed space-based observatories such as the JEM-EUSO mission~\cite{adams_2013} and radio detection instruments~\cite{romero-wolf_2013, motloch_2013, bray_2015} offer the possibility of significantly extending the sample of UHECRs available for source correlation analysis. 
% THIS IS SPECULATIVE: which is currently a perceived limitation of current ground based arrays. Alternatively, combining data from northern and southern ground arrays could improve the limited field of view of current ground-based observatories.

This paper aims to estimate the UHECR sample size that could lead to a statistically significant source catalogue correlation with a full-sky survey of UHECRs above the energy threshold of 60 EeV. Given current limitations on knowledge of source composition as well as galactic and intergalactic magnetic fields, we provide estimates for various assumptions on these parameters.
%WE DID NOT REALLY DEAL WITH THIS QUESTION AS IT IS STATED: We ask the question: given our current constraints on the galactic and intergalactic magnetic field and assuming an anisotropic source distribution, what are the future requirements in acceptance for a full-sky detector to measure source clustering in the sky with high significance? We provide a study of the expected degree of correlation given the strength and coherence length of galactic and intergalactic magnetic fields. 

There have been several simulation studies on the effects of cosmic ray scattering by galactic and intergalactic magnetic fields and their impact on source identification efforts~\cite{yoshiguchi_2003, sigl_2004, takami_2006, kashti_and_waxman_2008, koers_and_tinyakov_2009, takami_and_sato_2010, kalli_2011, rdeo_2014}. The approach presented here is not to provide another detailed simulation of cosmic ray propagation. Instead, we use existing simulation results to provide a parametric simulation that readily ties the behavior of composition, galactic and intergalactic magnetic field parameters to the catalogue correlation analysis presented by the Auger collaboration.

A related study explored the sensitivity of a JEM-EUSO-like instrument to detecting anisotropic arrival directions of UHECRs, assuming several different astrophysical source distributions, but a single magnetic field scattering model. They report that anisotropy would be observed in the lighter component of the UHECR population with a detector that has an order of magnitude increase in exposure over current experiments \cite{Oikonomou_2015}. Our work complements this approach by considering several magnetic field models, but with a single presumed source distribution. 

There have been other studies for all-sky instruments that employ different techniques for identifying anisotropy. One study employs the two-point correlation function to estimate the sensitivity of detecting multiplets, using the 2MRS catalogue~\cite{rdeo_2014}. Another study estimates the sensitivity to excesses of the dipole and quadrupole moments in the arrival direction distributions above an isotropic distribution~\cite{denton_2015}. These techniques do not employ the use of a source correlation catalogue to estimate the degree of anisotropy. In contrast, our work focuses on estimating the sensitivity to detecting a statistically significant correlation between UHECR arrival directions and a source catalogue, relative to an isotropic distribution. In the best possible case, the source catalogue and the correlation catalogue are identical.

% {\bf We need to expand this paragraph a bit. Are the works cited not able to provide a parametric simulation of GMF and IGMF parameters? We need to explain what the merits of this paper are. Merits of this paper: 1 We show the method to realize a complete correlation study with a given UHECR source catalogue. 2 We put everything together: igmf and gmf constraints, UHECR propagation (cosmological source selection and CMB interaction) and possible UHECR source catalogue}. There have been many simulations developed to study the scattering of UHECRs in the intergalactic magnetic field \cite{yoshiguchi_2003, sigl_2004, takami_2006, kashti_and_waxman_2008, koers_and_tinyakov_2009, takami_and_sato_2010, kalli_2011}. Our goal is not to develop a detailed cosmic ray propagation algorithm but rather to provide a parametric simulation that readily ties the behavior of galactic and intergalactic magnetic field parameters to the correlation of UHECR events to a parent source catalogue in a model-independent way.

The paper is organized as follows. In section 2 we cover UHECR scattering due to galactic and extragalactic magnetic fields. Section 3 presents a parametric model for cosmic ray propagation. Section 4 presents simulated correlation results of UHECRs with a source catalogue under different assumptions of galactic and extra-galactic magnetic fields.

%I made a change here. --Steph {\bf Figure} Figure showing constraints from gamma-rays, hubble expansion, CMB, and Faraday rotation measures.

\section{Cosmic Ray Scattering due to Galactic and Intergalactic Magnetic Fields}
The magnetic fields affecting cosmic ray deflection are not entirely understood. The intergalactic magnetic field (IGMF) has a wide range of uncertainty in its parameters given how difficult it is to observe its effects. The parameters that describe the IGMF are the magnetic field strength $B_0$ and coherence length $\lambda_B$. A long coherence length $\lambda_B$ means the magnetic field is constant in direction and magnitude over large distances, while a short $\lambda_B$ indicates a turbulent magnetic field that is varying direction and magnitude on small scales. Depending on the distance of propagation, cosmic ray scattering can depend as much on the magnetic field strength as it does on coherence length. 
The galactic magnetic field (GMF) has turbulent contributions from the disk and the halo, which are also poorly constrained. In the following we discuss the parameters of the intergalactic and galactic magnetic field and describe a parametric model of their scattering of ultra-high energy cosmic rays.

%For magnetic fields that are constant over the cosmic ray distance of propagation, $D$, the deflection is directly proportional to the distance to the source, while for magnetic fields with $\lambda_B$ smaller than the distance of propagation, the scattering is proportional to the geometric mean of the distance and $\lambda_B$.  % NO NEED TO DESCRIBE THE SCATTERING EQUATION IN THE INTRODUCTION.

\subsection{Intergalactic magnetic fields}
%\begin{itemize}
%\item Intro to our knowledge of them.
%\item What characterizes them.
%\item Current constraints.
%\item Parameters we consider
%\end{itemize}
Current bounds in the strength of the intergalactic magnetic field constrain $B_0$ to a range $10^{-17} - 10^{-9}$~G. The upper bound $B_0<10^{-9}$~G is due to the impact of intergalactic magnetic fields on cosmological perturbations and CMB anisotropies using Planck data~\cite{planck_2015}.
%The upper bound $B_0<10^{-9}$~G is due to the non-observation of Faraday rotation from the light of distant quasars~\cite{kronberg_1994} {\bf this is an outdated reference}. 
The lower bound $B_0>10^{-17}$~G is due to the non-observation of GeV $\gamma$-rays by the Fermi Large Area Telescope following from TeV $\gamma$-rays observed by HESS~\cite{neronov_2010}.

The intergalactic magnetic field coherence length $\lambda_B$ is also poorly constrained over a large range. A theoretical argument of magneto-hydrodynamic turbulence decay results in a constraint that $\lambda_B>0.1$~Mpc at $B_0=1$~nG and  $\lambda_B>10^{-6}$~Mpc at $B=10^{-15}$~G. A detailed review of constraints on the IGMF parameters can be found in~\cite{durrer_2013}.

A recent study~\cite{neronov_2013} reports evidence of a redshift dependence on the rotation measure from quasar light curves. The observations result in a measurement of intergalactic magnetic field strength $B_{0}=1^{+1}_{-0.3}$~nG. However, similar works did not find a strong correlation between rotation measure and redshift, due to the large variation in the intrinsic rotation measure of each source~\cite{xuhan_2014,banfield_2014,pshirkov_2015}. This means that $B_{0}$ could be well below 1~nG. As will be shown later, the IGMF, even at this strength, is not the dominant source of scattering.

Assuming a $B_{0}=1$~nG combined with the MHD constraints~\cite{durrer_2013}, we arrive at a lower bound on the coherence length $\lambda_B>$0.1~Mpc. It is interesting to note that while $\lambda_B=$~0.1 Mpc could be generated by AGN winds~\cite{bertone_2006, durrer_2013}, longer coherence lengths can currently only be explained by a cosmological origin of primordial magnetic fields~\cite{durrer_2013}.

\subsection{Galactic magnetic fields}
%\begin{itemize}
%\item Galactic magnetic fields have halo and disk component.
%\item Discuss parameters of disk component.
%\item Discuss parameters of halo component.
%\end{itemize}
The galactic magnetic fields, in addition to being described in terms of their strength $B_0$ and coherence length $\lambda_B$, are also bound over a distance $D$. The galactic magnetic fields are classified in terms of the disk and halo contribution, each with their own parameters $B_0$, $\lambda_B$, and $D$. The disk and halo magnetic fields can have regular ($ D\leq \lambda_B$) and turbulent ($D>\lambda_B$) components. 

The galactic disk's regular magnetic field effect on the scattering of UHECRs was studied by Stanev~\cite{stanev_1997} with a magnetic field strength of $B_0\sim$2~$\mu$G. The distance over which the UHECR is deflected is limited by the thickness of the disk and assumed to be $D\sim$2~kpc. The turbulent component of the galactic disk's magnetic field is assumed to have a strength of $B_{0}\sim$4~$\mu$G with a coherence length $\lambda_{B}\sim$50~pc~\cite{neronov_2009}. The magnetic field of the galactic halo is less well known. Studies conducted by Jansson et al. 2009,~\cite{jansson_2009} estimate a regular magnetic field with strength $B_{0}\sim$2~$\mu$G over a distance of $D\sim$8.7~kpc. Other measurements~\cite{sun_2008} indicate the halo magnetic field strength may be significantly higher. No observational constraints on the turbulent component of the galactic halo magnetic field are known. 
%See Table~\ref{tbl:Bfields} for a summary.

\subsection{Scattering of UHECRs by Magnetic Fields}
\label{sec:scat}
%\begin{itemize}
%\item Brief review of scattering models and their features.
%\item Motivate the model adopted for this work.
%\item Describe the model.
%\end{itemize}
%The overall sky distribution of the arrival directions of UHECRs below GZK limit (E$<60$x$10^{18}$eV [60 EeV]) seems to support the isotropy hypothesis \cite{naga_2000,abb_2004,molle_2008}.
%The potential anisotropy of super-GZK events, instead, has been regarded to provide an important clue that could unveil the sources of UHECRs, see \cite{ryu_2010ap}. 
%We want to study the effects of UHECR deflection by the IGMF on their apparent direction. 
%{\bf The first two paragraphs of section still needs work. It reads too much like a lab report.} For UHECR-IGMF scattering model in literature there are a lot of analytical models: \cite{waxman_1996}, \cite{lee_1995}, \cite{aha_2010} and \cite{act_1999}. 
Several parameterizations of the magnetic field scattering of UHECRs exist in the literature \cite{waxman_1996,act_1999,aha_2010}. 
%Waxman \& Miralda-Escud`e  consider cosmic rays energy $10$~EeV~$< E <40$~EeV, \cite{act_1999}, IGMF strength $B \lesssim 10^{-11}$G, and the coherence length $\lambda_{B}=1$Mpc and \cite{aha_2010} describes a model considering UHECRs only as protons, $Z=1$.
In this work, we adopt the results from Lee et al., 1995 ~\cite{lee_1995}, as presented in Neronov and Semikoz, 2009 ~\cite{neronov_2009}, which provide a parameterization valid for energies $E > 10$~EeV, including the varying scales for the coherence length $\lambda_{B}$, magnetic field strength $B_0$, and charge number $Z$.  This parameterization for the mean scattering angle of a UHECR due to interactions with a magnetic field is given by

\begin{equation} 
\vartheta_{scat}=2.6^{\circ}\left(\frac{E}{\mbox{100 EeV}}\right)^{-1}\left(\frac{D}{\mbox{50 Mpc}}\right)\left(\frac{B_0}{\mbox{$10^{-10}$ G}}\right)Z,
\label{eqn:scat1}
\end{equation} 
for a regular field and
\begin{equation} 
\vartheta_{scat}=0.23^{\circ}\left(\frac{E}{\mbox{100 EeV}}\right)^{-1}\left(\frac{D}{\mbox{50 Mpc}}\right)^{0.5}\left(\frac{B_0}{\mbox{$10^{-10}$ G}}\right)\left(\frac{\lambda_{B}}{\mbox{1 Mpc}}\right)^{0.5}Z
\label{eqn:scat2}
\end{equation}
for a turbulent field. 

UHECRs are scattered over a distance, $D$, which varies with magnetic field model. For intergalactic magnetic fields, $D$ is the distance to the source, while for galactic magnetic fields, it is the bounding distance relevant to the different regions of the galaxy. Here we average over the structure in the galactic halo and disk. A more detailed treatment is described in \cite{lee_1995}.

It is important to note that this parameterization is valid for small deflection angles ($\vartheta_{scat}\lesssim10^{\circ}$). We do not include a separate parameterization for higher scattering angles since it is not relevant to this study. As will be shown later in this work (see Sec. \ref{sec:source_corr_analysis}), only deflections well within the range of validity of this parameterization are shown to correlate to their sources.

%$E$ is the cosmic ray energy, $B$ is the magnetic field strength, and $Z$ is the atomic number. 
%In both cases the energy/charge attenuation distances toward the sources of the highest cosmic rays to be not larger than $D \sim 100$Mpc.
%The scattering angle $\vartheta_{scat}$ is the mean scattering angle of UHECRs propagating through a magnetic field.

%For larger scattering angles it does not make sense to correlating cosmic ray directions to candidate source positions. 
%In the case ($\lesssim10^{\circ}$), the energy spectrum of charged UHECRs from a given source is not significantly altered as compared to a straight-line propagation. However, if the sources cause an anisotropy in the UHECR flux, the directional correlation of ``hot spots" with possible sources will depend on the structure of the IGMF. 

%The scattering relations in Equations~\ref{eqn:scat1}~and~\ref{eqn:scat2} have 5 parameters ($E$, $B$, $D$, $\lambda_{B}$, $Z$). Based on the measurement of \cite{neronov_2013}, 

%The model given by Equations~\ref{eqn:scat1}~and~\ref{eqn:scat2} is also applied to the galactic magnetic field propagation. The components described as regular are assumed to have $\lambda_B\gg D$ while turbulent magnetic fields have $\lambda_B\lesssim D$. 
Table~\ref{tbl:Bfields} summarizes the magnetic field scales considered here and includes a representative scattering angle for a proton with energy $E=10^{20}$~eV. The energies used for this parameterization are the energy of the UHECR upon entering the magnetic field. For IGMF propagation, this is the energy of the UHECR at the source, while for galactic magnetic field scattering $E$ is the energy of the UHECR entering the galaxy.
\begin{table}[H]
\caption{Magnetic field contributions to UHECR scattering. The parameter $\vartheta_{p20}$ is for a proton with $E=10^{20}$~eV. The IGMF entry represents the smallest scattering angle possible with the parameters considered in this work.}
\label{tbl:Bfields}
\centering
\begin{tabular}{|l|l|l|l|l|}
\hline
Contribution & $B_{0}$ & $\lambda_{B}$ & $D$ & $\vartheta_{p20}$\\
\hline
Gal. Disk Reg. & 2 $\mu$G & n/a & 2 kpc   & 2.1$^{\circ}$\\
Gal. Disk Tur. & 4 $\mu$G & 50 pc & 2 kpc & 0.9$^{\circ}$\\
Gal. Halo Reg. & 2 $\mu$G & n/a & 8.7 kpc & 9.0$^{\circ}$\\
IGMF & 1 nG & $>0.1$ Mpc & $>$4 Mpc & $>$0.2$^{\circ}$\\
\hline
\end{tabular}
\end{table} 

\begin{figure}[htpb]
\centering
\includegraphics[width=0.75\linewidth]{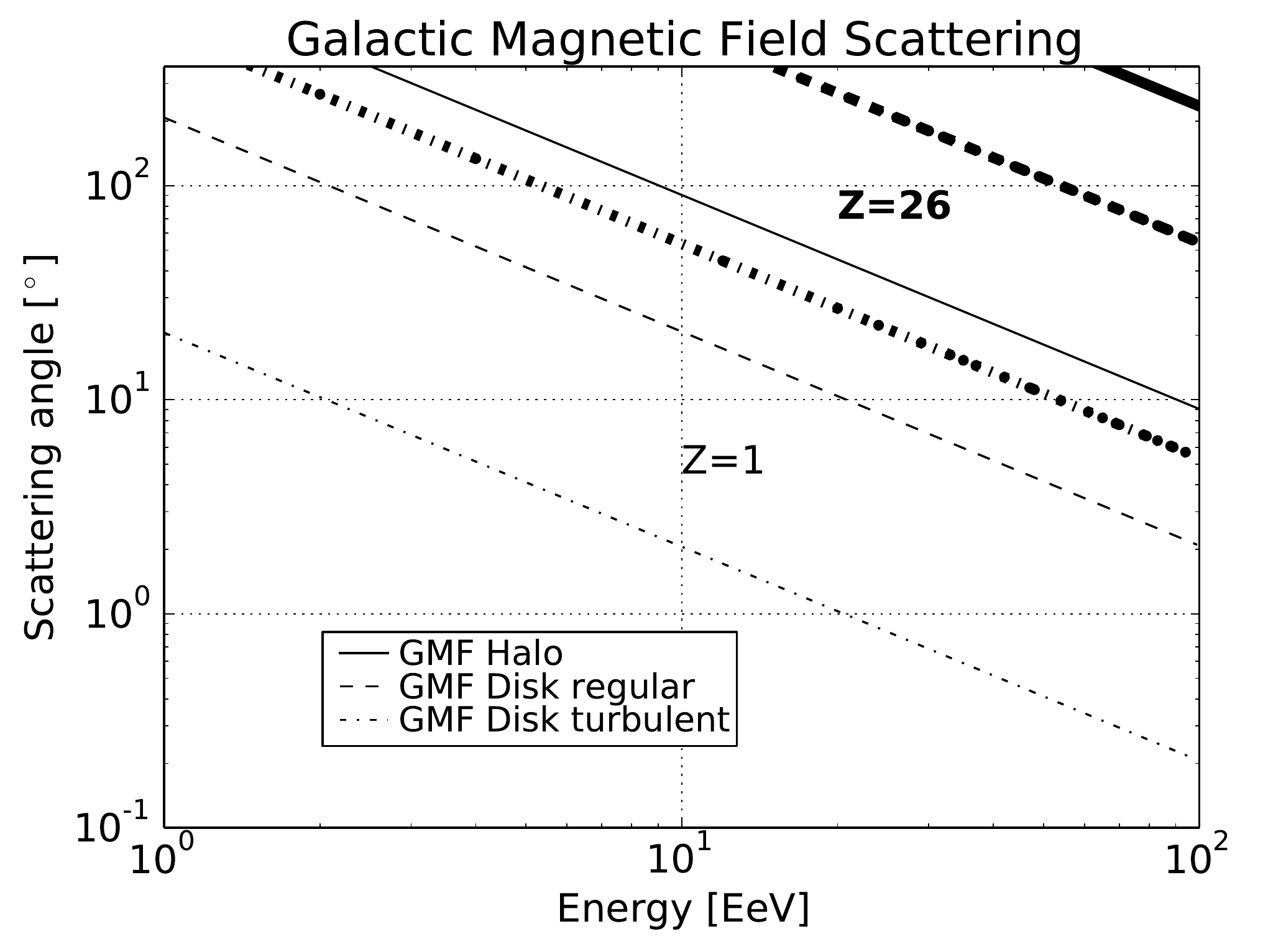} 
\caption{Mean scattering angle vs. cosmic-ray energy plots for different galactic magnetic field contributions (marked by line style). The thin lines are for protons while the bold lines are for iron nuclei. See Table~\ref{tbl:Bfields} for the parameters defining each galactic magnetic field contribution. See the text for a discussion of the range of validity of these parameterizations and how they are used in this study.}
\label{fig:gmf_scat}
\end{figure}

To give a sense of how the cosmic ray direction is affected by the galactic magnetic fields, Fig.~\ref{fig:gmf_scat} plots the mean scattering angle as function of the cosmic ray source energy for the halo regular, disk regular, and disk turbulent magnetic fields. The halo magnetic fields dominate the scattering while disk turbulent scattering tends to be a small effect. 
%The scattering is shown for both protons and iron nuclei. 
Iron nuclei are significantly deflected even for the galactic magnetic field disk turbulent contribution.

\begin{figure}[H]
\centering
\includegraphics[width=0.75\linewidth]{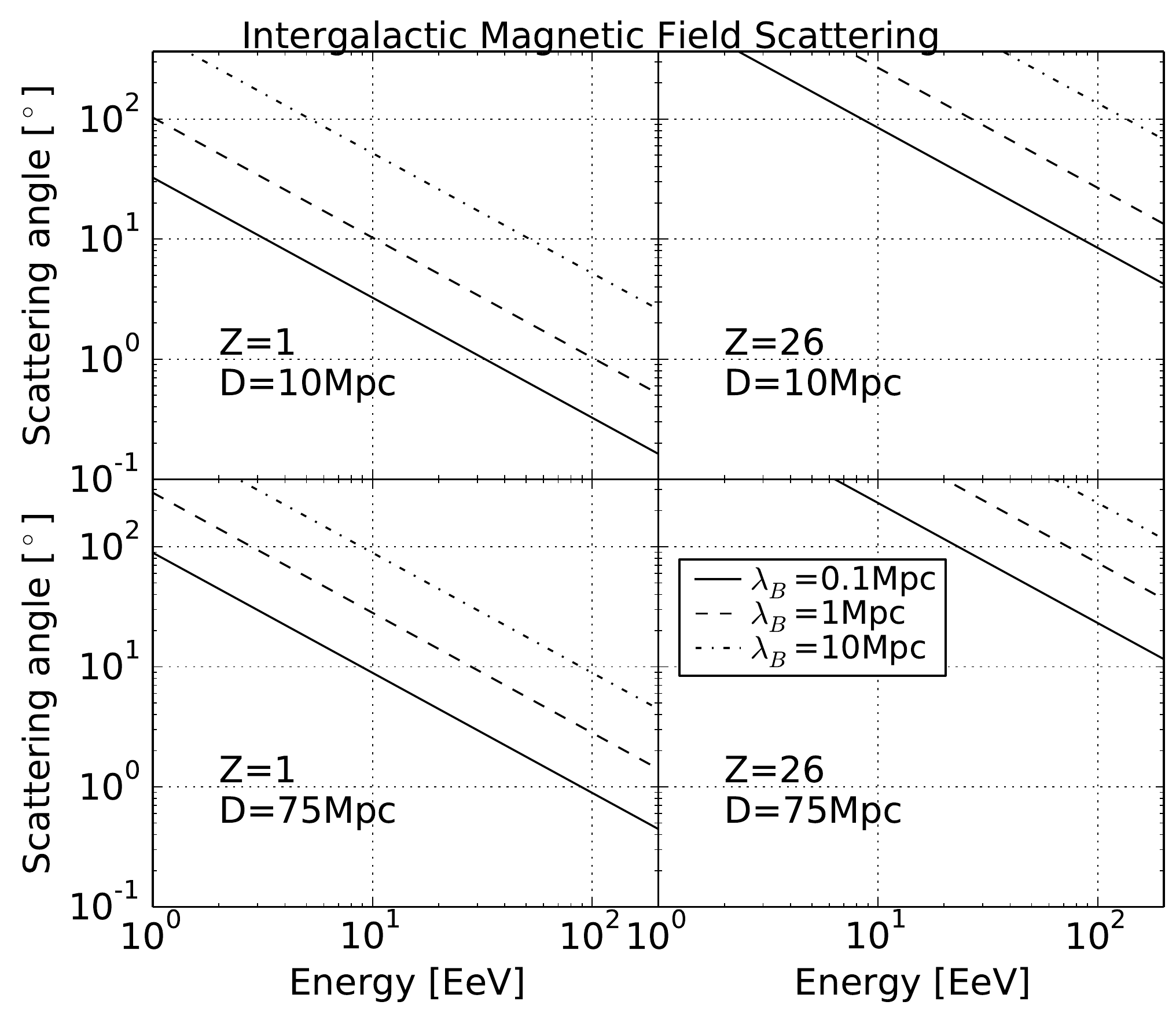} 
\caption{Mean scattering angle-energy plots for various combinations of magnetic field coherence length $\lambda_{B}$, cosmic-ray charge $Z$, and propagation distance $D$. See the text for a discussion of the range of validity of these parameterizations and how they are used in this study.}
\label{fig:igmf_scat}
\end{figure}

For the IGMF, we set $B_0=1$~nG and provide results for three coherence length values $\lambda_{B}$~=~0.1,~1,~10~Mpc according to the constraints provided by~\cite{durrer_2013}~and~\cite{neronov_2013}. Fig.~\ref{fig:igmf_scat} plots the mean scattering angle as function of the energy for different source distances: $10$~Mpc and $75$~Mpc for protons and iron nuclei. Iron nuclei display significant deflection even for relatively nearby sources.

%\subsection{Directional Scattering}
%From the Eq.$1$ \& Eq.$2$ we obtain the mean scattering angle. A description of the variance is not given in \cite{lee_1995}. We assume a Rayleigh distribution with mode $\sigma$ given by:

While the studies leading to the scattering models used here provide the average or root-mean-square scattering angles, they do not provide or discuss their statistical distribution.  For regular fields, the scattering angles are deflected systematically by the mean scattering angle given by Eqn. \ref{eqn:scat1}. For turbulent fields, we assume that the scattering angles (Eqn. \ref{eqn:scat2}) are Rayleigh distributed with  mode $\sigma_{scat}$, which is related to the mean scattering angle via
\begin{equation} 
\sigma_{scat}=\vartheta_{scat}\sqrt{\dfrac{2}{\pi}}.
\end{equation}
The scattering from a source catalogue is estimated by sampling this distribution.% In Fig.\ref{fig:rayleigh_example} we show a spherical representation in RA \& DEC of the cosmic ray events scattered around a source and the plot of the of the Rayleigh distribution of the scattering angle.

% \begin{figure}[H]
% \centering
% \includegraphics[width=1\linewidth]{figures/cenA.pdf} 
% \caption{Top: Points scattered in right ascension and declination around a simulated CenA source (201.37$^{\circ}$, -43$^{\circ}$). The Rayleigh distribution has $\sigma_{scat}$=3.1$^{\circ}$. Bottom: the distribution sampled event angular deviations from the source are fitted to a Rayleigh distribution (\textit{solid line}).}
% % \label{conf}
% \label{fig:rayleigh_example}
% \end{figure}

\section{Cosmic Ray Source Model and Propagation} 
\label{sec:scat}
%\begin{itemize}
%\item Brief review of cosmic ray propagation models. How is it typically done?
%\item Scattering during propagation. -- already covered in previous section.
%\item Interactions with cosmic photon backgrounds.
%\item Stepped energy loss.
%\end{itemize}
We model the sources of UHECRs as emitters following a power law energy spectrum with constant cosmic ray luminosity. The arrival flux and energy at Earth are affected by inverse-squared distance losses and attenuation due to interactions with cosmic photon backgrounds.
The following section describes the parametric model used to account for propagation energy losses.
%In this paper we adopt a parametric model to account for cosmic ray energy losses during propagation.
%By using this model we include an energy/distance weighted cosmic ray spectrum from our source catalogue.

%Now we describe the technique to simulate a sample of UHECRs from AGNs.  

\subsection{Cosmic Ray Source Model}
\label{sec:src_model}
%\begin{itemize}
%\item What we assume for the cosmic ray source model and its flux.
%\item How this translates to the flux on Earth
%\end{itemize}
We assume the cosmic ray source luminosity spectrum $L_{src}(E_{src})$ (the rate of particles emitted at a given energy) for source energy $E_{src}$ for each source in the catalogue follows a power law with index $\gamma_g$
\begin{equation}
L_{src}(E_{src})=L_{0}\left(E_{src}/E_{0}\right)^{-\gamma_g}
\end{equation}
where $L_{0}$ is assumed to be a universal source luminosity at a reference energy of $E_{0}=10^{19}$~eV. We consider values of $\gamma_g=2.0$ and $\gamma_g=2.7$ but present results only for the latter. We found no significant difference in our results between the two values of $\gamma_g$ considered.
%Following~\cite{berezinsky_2006}, we consider values of $\gamma_g=2.0$ and $\gamma_g=2.7$.

We assume a detector with effective area $A_{eff,det}$ and exposure time $T$ that is the same for each source indexed by $i$. The total number of particles with an arrival energy at the observation point $E_{obs}$ that is above a cut energy of $E_{cut}$ is given by
\begin{equation}
N(E_{obs}>E_{cut})=T A_{eff,det}\sum_{i=1}^{M}\int_{0}^{\infty}dE_{src} \Theta(E_{obs}(E_{src})-E_{cut})\frac{L_{src}(E_{src})}{(1+z_{i})4\pi d^2_C(z_i)}.
\label{eqn:num_events}
\end{equation}
In this equation $M$ is the number of sources in the catalogue. The function $E_{obs}(E_{src})$ is the arrival energy resulting from a particle with energy $E_{src}$ at the source at redshift $z_{i}$. The propagation is described in the subsections below. $d_C(z_i)$ is the comoving distance of the source. The function $\Theta(E_{obs}(E_{src})-E_{cut})$ is the Heaviside step function requiring that the energy of arrival to the observer be greater than the cut value. See Appendix A for a derivation of Equation~\ref{eqn:num_events}.

For this study, we sample the catalogue to obtain the desired number of events assuming the detector has necessary exposure. The sampling is performed following the sum and integral in Equation~\ref{eqn:num_events}. Each source in the catalogue is sampled according to $[(1+z_i)d^{2}_C(z_i)]^{-1}$, which in practice is achieved by rejection sampling.  The source energy $E_{src}$ is sampled according to the inverse power law with index $\gamma_g$, and propagated to the energy of arrival to the observer $E_{obs}$. If $E_{obs}>E_{cut}$, the event is accepted, otherwise it is rejected.

%The cosmic ray flux on Earth $F_{Earth}$ is given by
%\begin{equation}
%F_{Earth}(E)=\int_{0}^{\infty}dE' \ Q(E,E',z)\frac{L_{src}(E')}{4\pi d_C^2}
%\end{equation}
%where $E'$ is the energy of the particle at the source and $E$ is its energy upon arrival, $z$ is the source redshift, and $d_C$ is the comoving distance between the source and Earth. $Q(E,E',z)$ is a transfer function that accounts for the energy loss due to propagation of a source at redshift $z$ to Earth through the cosmic photon backgrounds. These effects are described below.

%In this paper we will consider both a pure proton composition and a pure iron composition to characterize the two extremes.

%\subsection{Cosmic Ray Energy Propagation}
\subsection{Cosmic Ray Energy Loss Length}
\label{sec:loss_length}
\begin{figure}[H]
\centering
\includegraphics[width=0.75\linewidth]{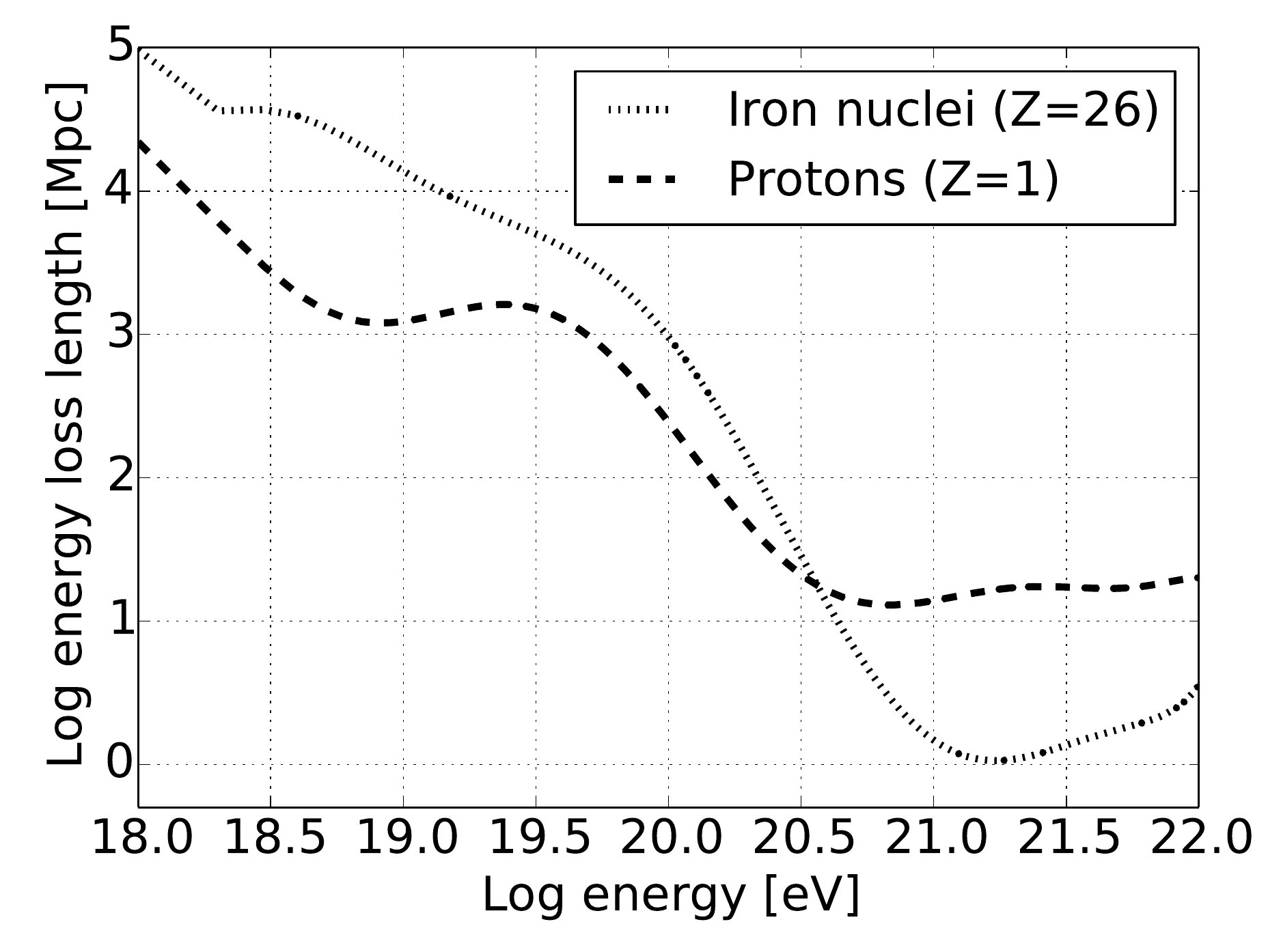}
\caption{Energy loss length as function of cosmic ray energy. This plot has been implemented by considering \cite{kotera_2011} for protons ($Z=1$) and \cite{stecker_1999} for iron nuclei ($Z=26$).}
\label{fig:loss}
\end{figure}

Cosmic ray energies are attenuated through particle interactions as they propagate through the Universe. Protons at the highest energies lose energy by producing pions through interactions with the CMB \cite{greisen_1966, zatsepin_and_kuzmin_1966}. The energy loss length for protons, shown here in Fig.~\ref{fig:loss}, shows that UHECRs with energies above $6\times10^{19}$eV are likely to be substantially attenuated. We adopt the proton loss length calculations by \cite{kotera_2011}, based on the analytical formulae from \cite{stecker_1968}.

Heavy nuclei, such as iron, lose energy through pair production on the CMB and photo-disintegration through interactions with CMB and IR-UV background photons \cite{puget_1976}.  Photohadronic energy loss has been re-examined in recent years, as they are based on empirical measurements of the intergalactic background radiation and photonuclear interactions.  We use \cite{stecker_1999} as our model for iron energy loss length, which is representative of the modern calculations \cite{epele_1998, bertone_2002, khan_2005}.

\subsection{Cosmic Ray Energy Propagation}
As the cosmic ray propagates through space, it is subject to adiabatic losses due to the expansion of the Universe as well as losses due to interaction with background photons. The density and energy of background photons is also changing as the Universe expands. To account for these losses, we discretely propagate the cosmic ray energy from its source at redshift $z$ to $z=0$ in steps of $\Delta z$. Given the energy $E_{j}$ at step $j$, the energy $E_{j+1}$ at step $j+1$ is given by 
\begin{equation}
E_{j+1}=E_{j}\left(1 - \frac{\Delta z}{1+z_j} - \frac{\Delta z}{\lambda_{\gamma}(E_j)}\frac{ cH^3_0}{H^2(z_j)}  \right)
%-\frac{\Delta z H^3_0}{H^2(z_k)}\frac{c}{\lambda_{\gamma}(E_k)}
\label{eqn:discrete_eprop}
\end{equation}
The second term in the right hand side is due to the adiabatic redshift losses while the third term is given by the interaction with background photons corrected for their evolution as the Universe expands. The energy-dependent cosmic ray attenuation length due to photon interactions, discussed in Section~\ref{sec:loss_length}, is given by $\lambda_{\gamma}(E)$. The Hubble constant today is given by $H_0$ while the Hubble parameter at redshift $z$ is given by $H(z)$. Equation~\ref{eqn:discrete_eprop} is derived in Appendix B.
%\begin{equation}
%E_{k+1}=E_{k} \exp\left[-{\dfrac{D_{k}}{L(E_{k})}}\right],
%\end{equation}
%where the energy of the \textit{k+1}-step is equal to the energy multiplied for an exponential function that depends on the source distance ($D$) at the \textit{k}-step and on the loss length ($L$) at the \textit{k}-step. 

\begin{figure}[H]
\centering
\includegraphics[width=0.75\linewidth]{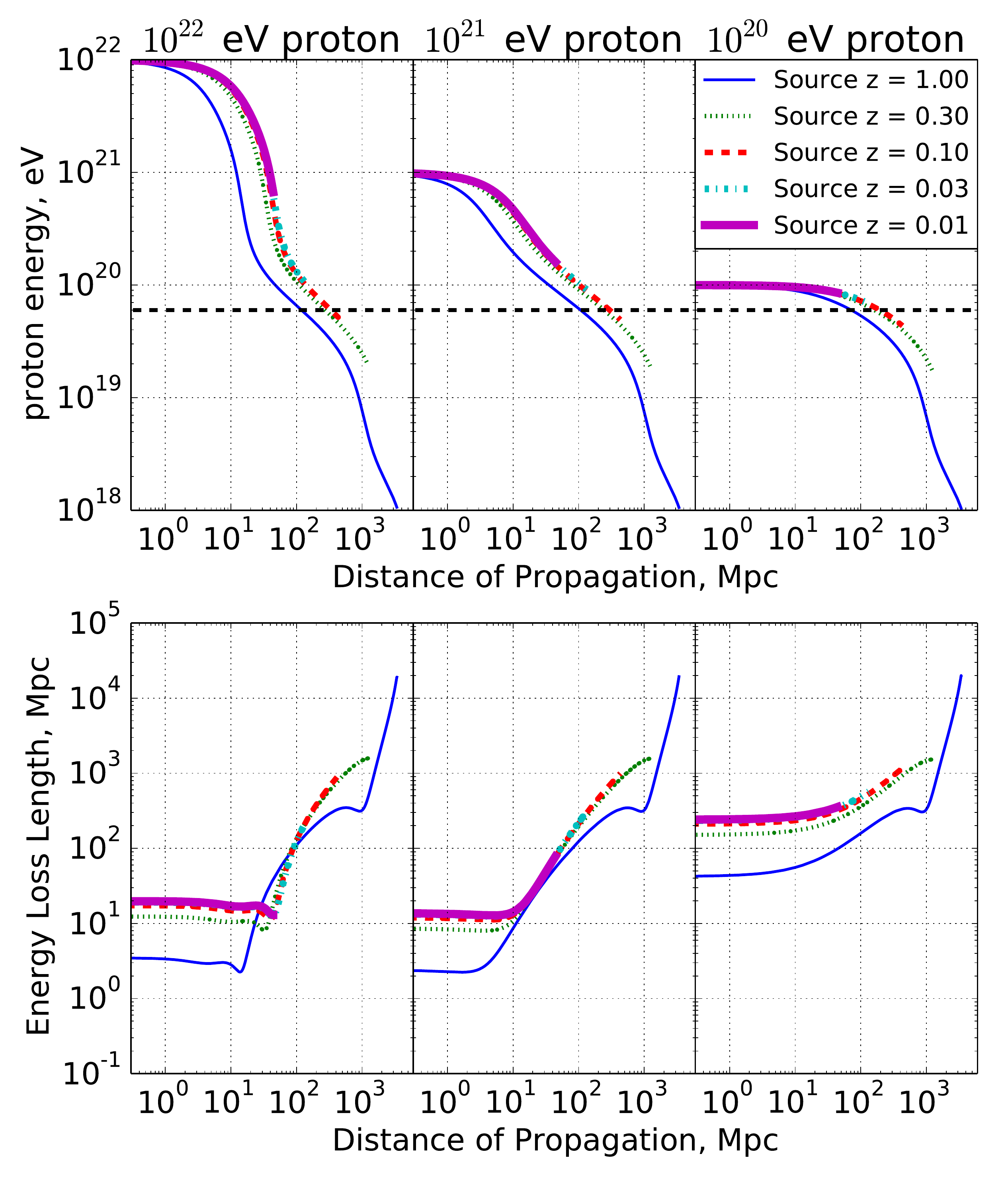} 
\caption{Examples of cosmic ray proton propagation for a source at redshift $z$=0.01, 0.03, 0.1, 0.3, and 1.0 at energies $E=10^{22}, 10^{21},$ and $10^{20}$~eV. The top panels show the proton energy evolution as a function of comoving distance of propagation. The bottom panel shows the energy loss length as a function of distance. This varies depending on the source redshift due to the changes in background photon density.}
\label{fig:proton_prop}
\end{figure}

\begin{figure}[H]
\centering
\includegraphics[width=0.751\linewidth]{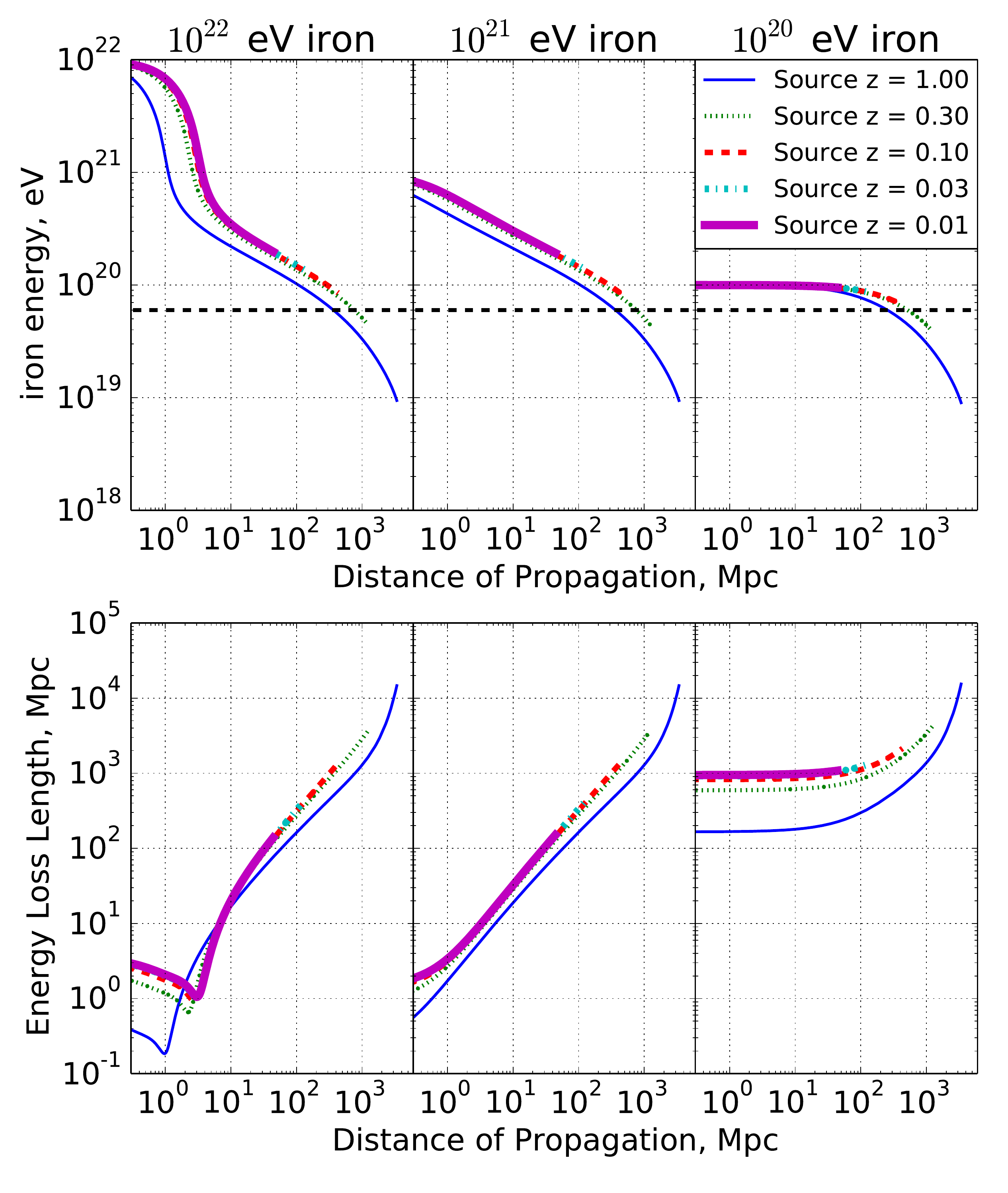} 
\caption{Examples of cosmic ray iron nuclei propagating for a source at redshift $z$=0.01, 0.03, 0.1, 0.3, and 1.0 at energies $E=10^{22}, 10^{21},$ and $10^{20}$~eV.}
\label{fig:iron_prop}
\end{figure}

Figures \ref{fig:proton_prop} and \ref{fig:iron_prop} plot the cosmic-ray energy and energy loss length as a function of propagation distance assuming sources located at various redshifts $z$. Assuming the energy loss length models for protons and iron nuclei in Figure \ref{fig:loss}, it is evident that for sources with $z>0.03$ cosmic rays reach Earth with energies below $6\times10^{19}$~eV. Cosmic rays observed above this $6\times10^{19}$~eV must originate from nearby sources. 

Figures \ref{fig:proton_prop} and \ref{fig:iron_prop} demonstrate that the energy loss length throughout the trajectory of a cosmic ray can vary significantly depending on its source redshift. This occurs because the background photon density and energy distribution evolve with redshift (see Appendix B), which also means that the energy loss length (Fig. \ref{fig:loss}) evolves with redshift. For example, at source redshift $z=0.3$, the photon density and mean energy at the beginning of the cosmic ray trajectory are lower than for a particle at source redshift $z=1.0$, resulting in a longer energy loss length. As the cosmic rays propagate from sources at different redshifts, they will lose energy at different rates. The photon density and energy distribution along their trajectories also change at different rates for each case.

%Note that the energy loss length varies with propagation distance as the background photon density changes, getting longer at 
% The step size we use $\Delta z = 10^{-5}$ as it does not affect the slope of the loss length function significantly ($< 10\%$). 
% The loss length is a function of energy. To describe this relation we adopt the model present in \cite{stanev_2004} for $Z=1$. 
%This model provides the variation of the energy loss length as function of the cosmic ray energy in the assumption of $1$~nG magnetic fields for proton. 
%At energies below $100$EeV the proton energy loss length is definitely longer than that of gamma rays. %At energies above $500$EeV the difference is only a factor of 2, with very small energy dependence. 
%The general conclusion from that analysis of the energy loss of protons and gamma rays in their propagation through the Universe is these ultra high energetic particles cannot survive at distances of more than few tens of Mpc and sources of the detected cosmic rays have to be located in our cosmological neighborhood. 

\begin{figure}
        \centering
        \begin{subfigure}[H]{0.75\textwidth}
                \includegraphics[width=\textwidth]{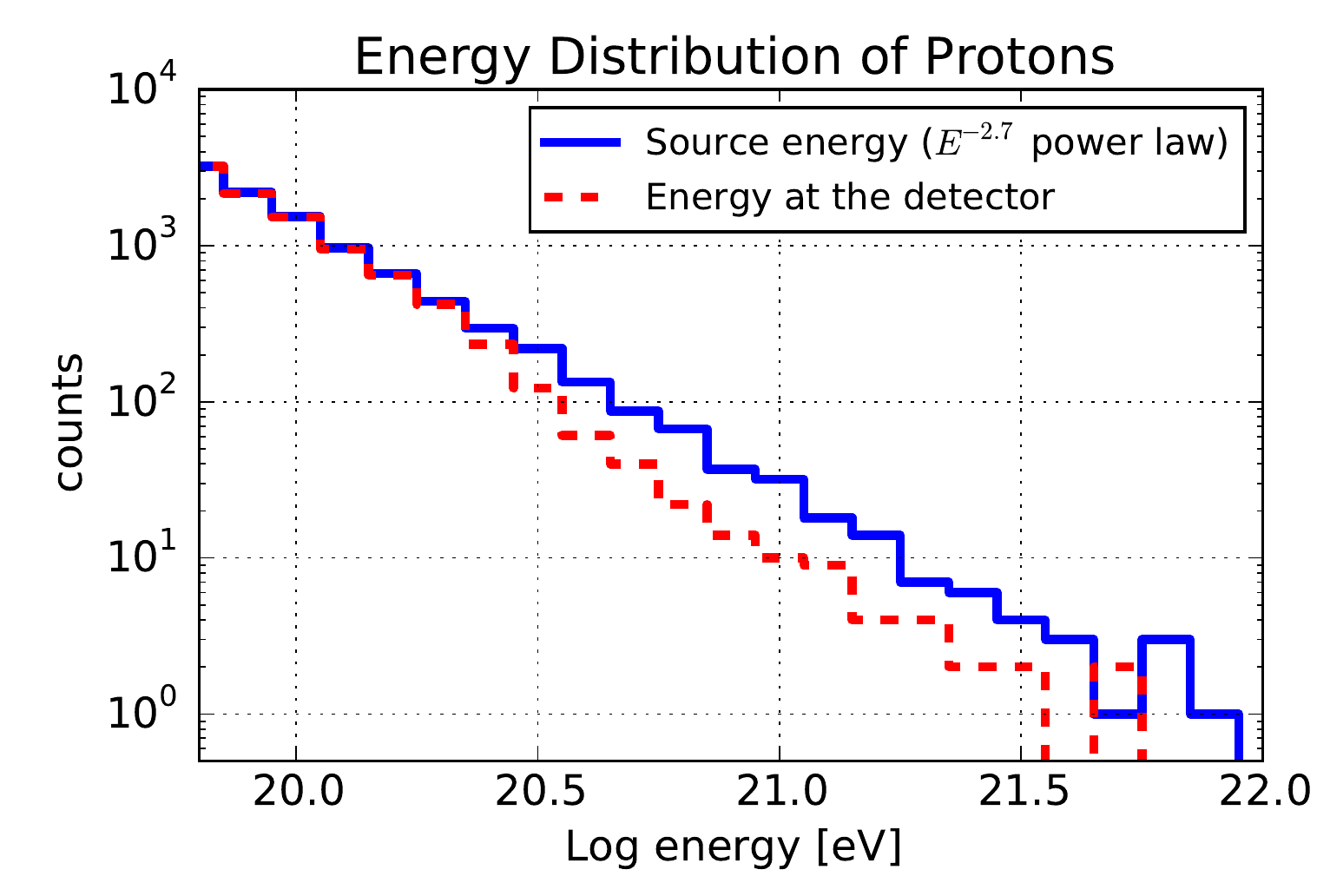}
                %\caption{}
                \label{bh}
        \end{subfigure}
        ~ 
        \begin{subfigure}[H]{0.75\textwidth}
                \includegraphics[width=\textwidth]{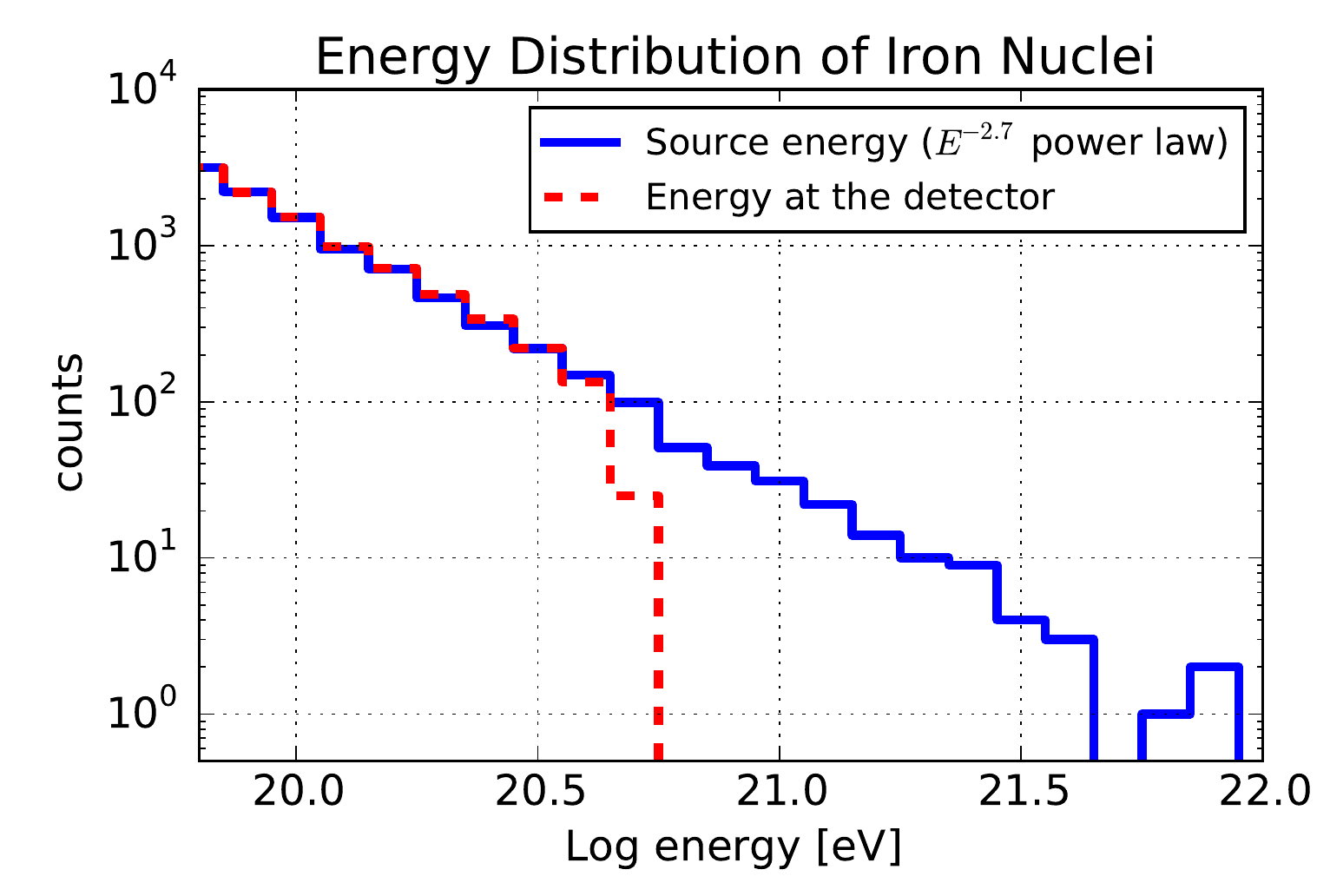}
                %\caption{}
                \label{bh1}
        \end{subfigure}
        \caption{Energy distribution of cosmic rays propagated to Earth for 10,000 samples of sources sampled according to the VCV catalogue distance distribution assuming an $E^{-2.7}$ luminosity energy spectrum. (Top) Proton UHECR distribution energy at the source and at the Earth. (Bottom) Iron UHECR distribution energy at the source and at the Earth.}\label{fig:energy_distrib}
\end{figure}

In Fig.~\ref{fig:energy_distrib} we show a simulated distribution of cosmic ray energies at Earth obtained from sampling 10,000 events from an $\gamma_g = -2.7$ source luminosity spectrum. The distances have been sampled from the VCV catalogue assuming all sources have the same luminosity (see next Section for details on sampling). The distribution of energies for iron nuclei shows a dramatic cutoff for energies $>5\times 10^{20}$~eV compared to protons. The location of the cutoff is due to the dominant contribution of nearby sources in the VCV catalogue, the nearest being at $\sim$4~Mpc.
% A small pile-up of events appears for energies for iron nuclei in the energy range between$3\times10^{20}$ and $6\times10^{20}$. 
% The the distribution of the proton UHECR energies at the source and at the detector for a catalogue sample. Similarly, in Fig.\ref{bh} we can observe the UHECR iron distribution energy at the source and at the detector, by following the model presented in \cite{berezinsky_2006}. Then we propagate these cosmic ray energies from the sources up to the detector using Eq.$1$. During this propagation UHECR energy interacts with CMB. All the energies are shifted to lower values, see Fig.\ref{bh1}. From this distribution we collect a sample by considering a cut-off at $60$EeV (GZK limit).   

\section{Correlation to the VCV source catalogues}
%\begin{itemize}
%\item VCV source catalogue description and motivation.
%% \item Characterize VCV catalogue before and after the squared distance propation.
%% \item Source correlation analysis.
%% \item Results.
%\end{itemize}
Two approaches may be used in looking for correlations among known sources and UHECR arrival directions.  One is to use a complete, uniform survey of matter in the universe. High correlations among such maps would indicate that UHECR production follows from regions of high star formation and gas density. The second approach is to look for correlations with catalogues of candidate UHECR acceleration sites, such as active galactic nuclei (AGN) or gamma ray bursts (GRBs).

The 2MASS survey \cite{2mass} and the PSCz surveys \cite{pscz} use infrared and near infrared observations to make the most uniform maps available. Dust in older, dimmer galaxies and proto-galaxies alike emit in the infrared, so they are the best tracers of large scale structure.  Similarly, Berlind and others \cite{berlind2011} have generated simulation catalogues based almost entirely on the matter density and $\Lambda$CDM N-body simulations.

The Veron-Cetty \& Veron (VCV) catalogue 12th edition \cite{vcv} is a compendium of known AGN, largely derived from the 2dF catalogue and the Sloan Digital Sky Survey. While it is known to be a non-uniform survey, we use it here based on the precedent set by prior searches for correlations with AGN by Auger \cite{abraham_2007, abreu_2010, aab_2015}, HiRes \cite{abbasi_2008}, and TA \cite{abuzayyad_2012}.  However, hereafter, we treat it as a mock catalogue of UHECR sources, by using it both as the source distribution and catalogue for correlation analysis. Our objective is not to test whether UHECRs are produced by AGN in the VCV catalogue, but rather to characterize the ability to correlate given the effects of scattering and energy losses. 

\subsection{Characterization of the VCV Catalogue}
%\begin{itemize}
%% \item Why VCV?
%\item Characterize VCV catalogue before and after the squared distance propagation.
%%\item Source correlation analysis.
%\item Distance distribution.
%\item Statistical angular separation between VCV sources.
%\end{itemize}
Treating the VCV catalogue AGN as the sources of ultra-high energy cosmic rays, we characterize the expected angular separations between sources. The goal is to estimate the angular uncertainties required to possibly distinguish between catalogue sources.
In this work, we will assume an optimistic angular resolution of $1^{\circ}$, given that orbiting fluorescence detectors such as JEM-EUSO~\cite{adams_2013, biktemerova_2015} expect angular resolutions between $1^{\circ}$ and $3^{\circ}$ and that radio-detection satellites \cite{romero-wolf_2013, motloch_2013} expect angular resolutions of $\sim1^{\circ}$. We model the detector angular resolution by sampling a Rayleigh distribution around the arrival direction.

\begin{figure}[H]
\centering
\includegraphics[width=0.75\linewidth]{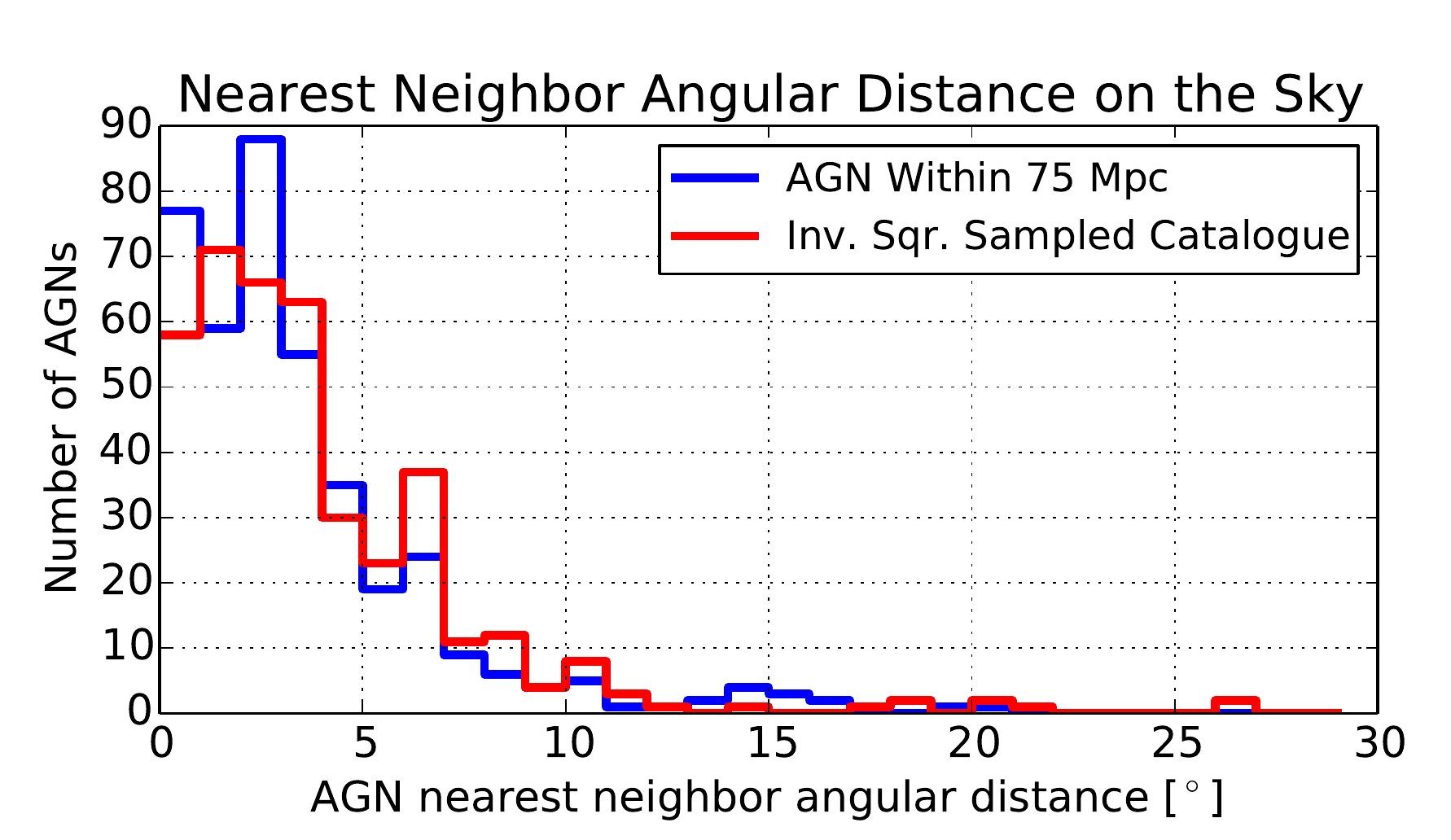} 
\caption{Veron-Cetty \& Veron catalogue 12th edition AGN nearest neighbor angular distance before and after inverse-square distance sampling.}\label{cluster}
\end{figure}

The ability to distinguish between VCV catalogue AGN can be characterized by looking at the nearest neighbor angular distance of each AGN. In Fig.~\ref{cluster} we show the distribution of AGN nearest neighbor angular distance for both the catalogue with sources $\leq$75~Mpc and the inverse-square distance sampled version. Both distributions are similar and we find that $\sim$1/3 of sources have a neighbor within $2^{\circ}$. Within an angular distance of $5^{\circ}$, 70\%-80\% of sources have a neighbor in the VCV catalogue. This level of source confusion would make it difficult to identify individual AGN catalogue sources with high confidence, regardless of magnetic field scattering effects. 
% If the detector a resolution $\leq 2^{\circ}$ it will be able to discriminate most of the AGNs to their nearest ones. In this work we will adopt an angular resolution value of $2^{\circ}$.

\begin{figure}[H]
\centering
\includegraphics[width=0.75\linewidth]{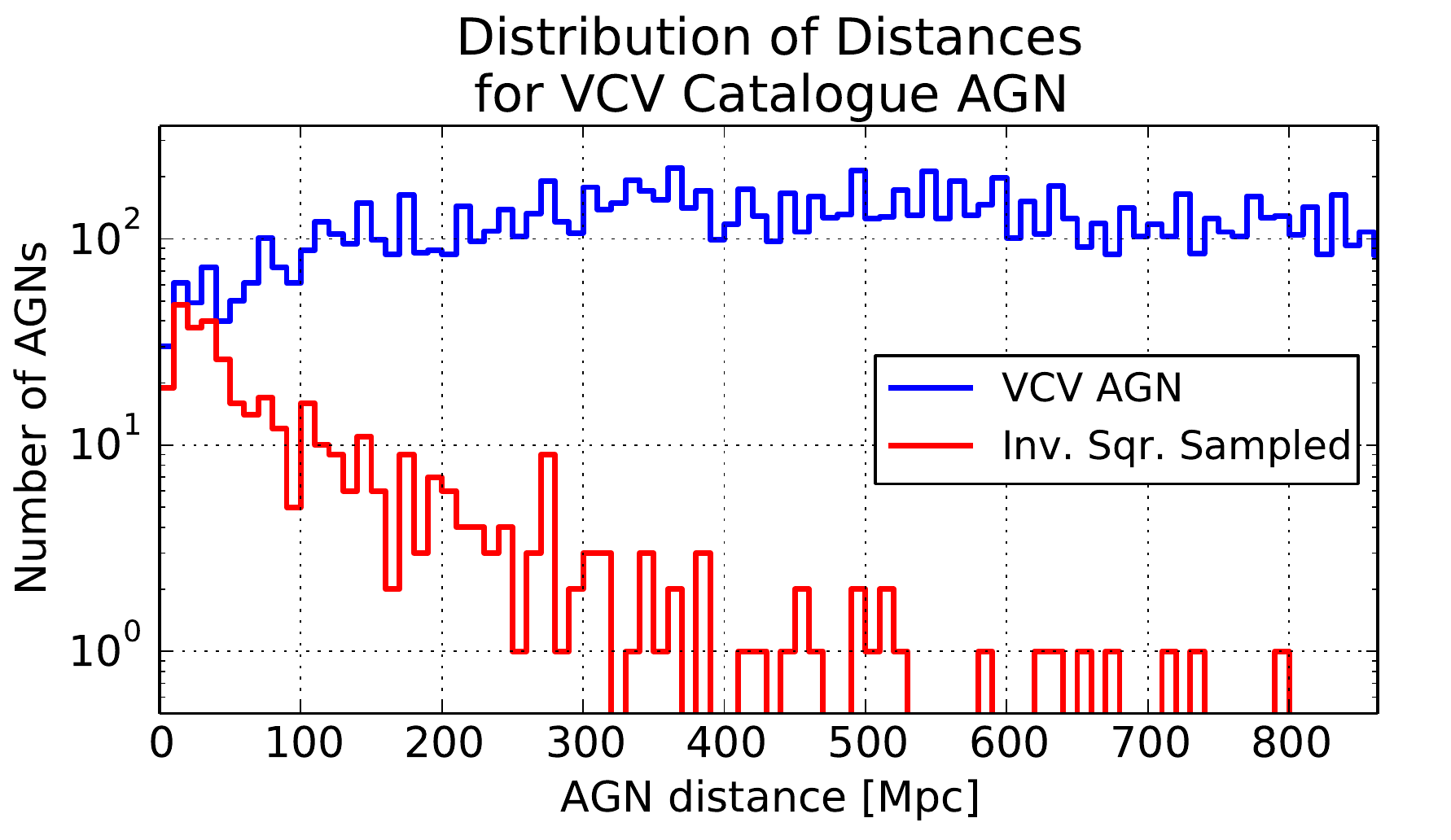} 
\caption{Veron-Cetty \& Veron catalogue 12th edition AGN distance distribution before and after inverse-square distance sampling of the catalogue. For this figure, a source is only allowed to be sampled once. Without this requirement (as done in the simulations performed here), the contribution due to the nearest sources significantly outweighs the rest.}\label{fig:inv_sq_sampling_distance}
\end{figure}

Another important effect to take into account, assuming the source AGN in the catalogue have similar luminosities, is the inverse-square distance source rate reduction described in Section~\ref{sec:src_model}. A distant source will contribute a lower flux on Earth compared to a nearby one simply due to the flux-luminosity relationship.
% In Fig. \ref{VCV_sources} we plot, in galactic coordinates ($l$,$b$), the AGNs present in Veron-Cetty \& Veron (VCV) catalogue 12th edition \cite{vcv} with a distance $D \leqslant 75$Mpc ($z \leqslant 0.018$), as proposed by the Auger collaboration \cite{auger_2007}. 
% In Fig. \ref{VCV_source_afterpropagation} we plot in galactic coordinates ($l$,$b$) the elements of the catalogue VCV 12th edition according to inverse square distance weighting. The distribution of AGN is, overall, similar. 
It can be seen in Fig.~\ref{fig:inv_sq_sampling_distance} that the inverse-square distance sampling of VCV catalogue sources alone favors nearby sources. Despite the increased abundance of far away sources, the probability density that sources at a distance $>$100~Mpc contribute to the arrival flux on Earth is reduced by more than an order of magnitude and becomes negligible past 300~Mpc.

It is worth mentioning that the number distribution of AGN as a function of distance in the VCV catalogue is not representative of the source evolution of AGN within a few hundred megaparsecs from Earth due to selection effects. The VCV sources, used as a source catalogue, would correspond to a strong negative redshift evolution that is unlike any ultra-high energy cosmic ray source candidate population. This approach leads to an increased contribution from nearby sources, with reduced intergalactic magnetic field scattering, compared to more distant sources. The lower bound resulting from this approach is therefore likely optimistic compared to using a source redshift evolution of a complete AGN catalogue. To avoid the complexities associated with densifying the VCV catalogue we use its listed sources as they are published keeping in mind that, although perhaps optimistic, this approach results in a perfectly valid lower bound. 

% In Fig.\ref{dist} we plot the AGN distance distribution of this catalogue. We notice that $\sim 46\%$ of the galaxies populate the range $50-75$Mpc. This means that almost half of the galaxies in the VCV catalogue will be excluded through the inverse distance square weighting of the fluxes very often and so a most cosmic ray events come from a small sample of AGNs with a distance $D \lesssim 20$Mpc, as shown in Fig.\ref{dist1}. 

\subsection{Source Correlation Analysis} \label{sec:source_corr_analysis}
%\begin{itemize}
%\item Follow the prescription laid out by Auger.
%\item Simulate isotropic arrival distribution to determine source correlation significance.
%\item Vary the correlation angle and source distance cutoff to characterize behavior of correlations.
%\item Look at including Galactic magnetic (disk and halo, disk only) or just IGMF.
%\item Example plots (let's keep this under control, use the preferred scenarios. All fields on, IGMF only. Proton vs. Iron).
%\end{itemize}
In this section, we describe the correlation of cosmic rays sampled from the VCV catalogue to VCV catalogue sources. The goal is to determine under which scenarios it is possible to discriminate between a catalogue correlation and an isotropic source distribution with $>5\sigma$ confidence as a function of cosmic ray sample size.

The correlation analysis follows the procedure applied by the Auger collaboration \cite{abraham_2007, abreu_2010}. %Starting with the source catalogue, a distance cutoff is applied for the correlation. 
We compare the arrival direction of an event with the position of the source in the catalogue. An event correlates to the catalogue if the angular distance between the source and the arrival direction of the cosmic ray is within a correlation angle, $\psi$. We will explore the effect of varying this parameter in our simulations. Given a sample of $N$ cosmic ray events, $k$ of which correlate to the catalogue, the probability of the data being correlated to the catalogue is given by that fraction $p_{data}=k/N$. We apply a cut on the distance of sources used for correlating against the event sample. We call this the correlation distance cutoff. Based on a data-driven optimization study, the Auger collaboration \cite{abraham_2007, abreu_2010} fixed the correlation angle at $\psi=3.1^{\circ}$ and the correlation distance cutoff to 75~Mpc. The correlation distance cut restricts the sources used in the correlation analysis to 50\% of the total number of sources in the VCV catalogue.

The confidence interval for $p_{data}$ is estimated from a binomial distribution $P(p_{data})=C_{b}(k,N)p_{data}^{k}(1-p_{data})^{N-k}$, where $C_{b}(k,N)$ is the binomial coefficient. We calculate $p_{data}$ for both a simulated arrival direction data set generated from the source catalogue and from an isotropic distribution. The isotropic distribution randomizes the arrival directions according to a uniform spherical distribution. We calculate the difference,  $\Delta p_{data}$, between $p_{data}$ for the catalogue-generated distribution and for the isotropic distribution. The statistical significance from isotropy is then calculated as the number of confidence intervals between $p_{data}$ for the catalogue-generated distribution and for the isotropic distribution. We express the statistical significance in units of $\sigma$, the Gaussian-distribution-equivalent number of standard deviations.

The parameters tested in this study are source composition and the properties of the galactic and intergalactic magnetic fields. We look at the effects of galactic and intergalactic magnetic fields by testing the following scenarios:
\begin{itemize}
\item GMF halo and disk (regular+turbulent) components and IGMF.
\item GMF disk (regular+turbulent) component and IGMF.
\item IGMF only.
\end{itemize}
in addition to testing the IGMF $\lambda_B$ at 0.1, 1.0, and 10 Mpc for a magnetic field strength of $B_0$=1 nG in each case. We also compare source compositions comprising 100\% protons and 100\% iron nuclei.

% In these three scenarios we vary the IGMF coherence length $\lambda_B= 0.1$, $1$, $10$ Mpc. We take into consideration the number of events we consider in our simulation. we run our analysis for $10$, $100$, $1000$ UHECRs events, to understand which is the event rate we need in order at least to discriminate anisotropic UHECR distribution from the isotropic one.

Rather than describing in detail the results of each scenario, we will first give an example of the analysis for source correlation analysis of 1000 proton events with arrival energies greater than 60 EeV detected with an angular resolution of 1 degree, assuming that the protons are scattered in both the GMF halo and disk. All of the scenarios considered here assume a 100\% detector efficiency with respect to energy and perfect energy resolution. Our results place a lower bound on the sensitivity to source catalogue correlations. Results with other simulation parameter values of interest are shown in Tables \ref{table:sigmas_1deg} and \ref{table:sigmas_3deg}.

\begin{figure}[htbp]
        \centering
                \includegraphics[width=0.75\textwidth]{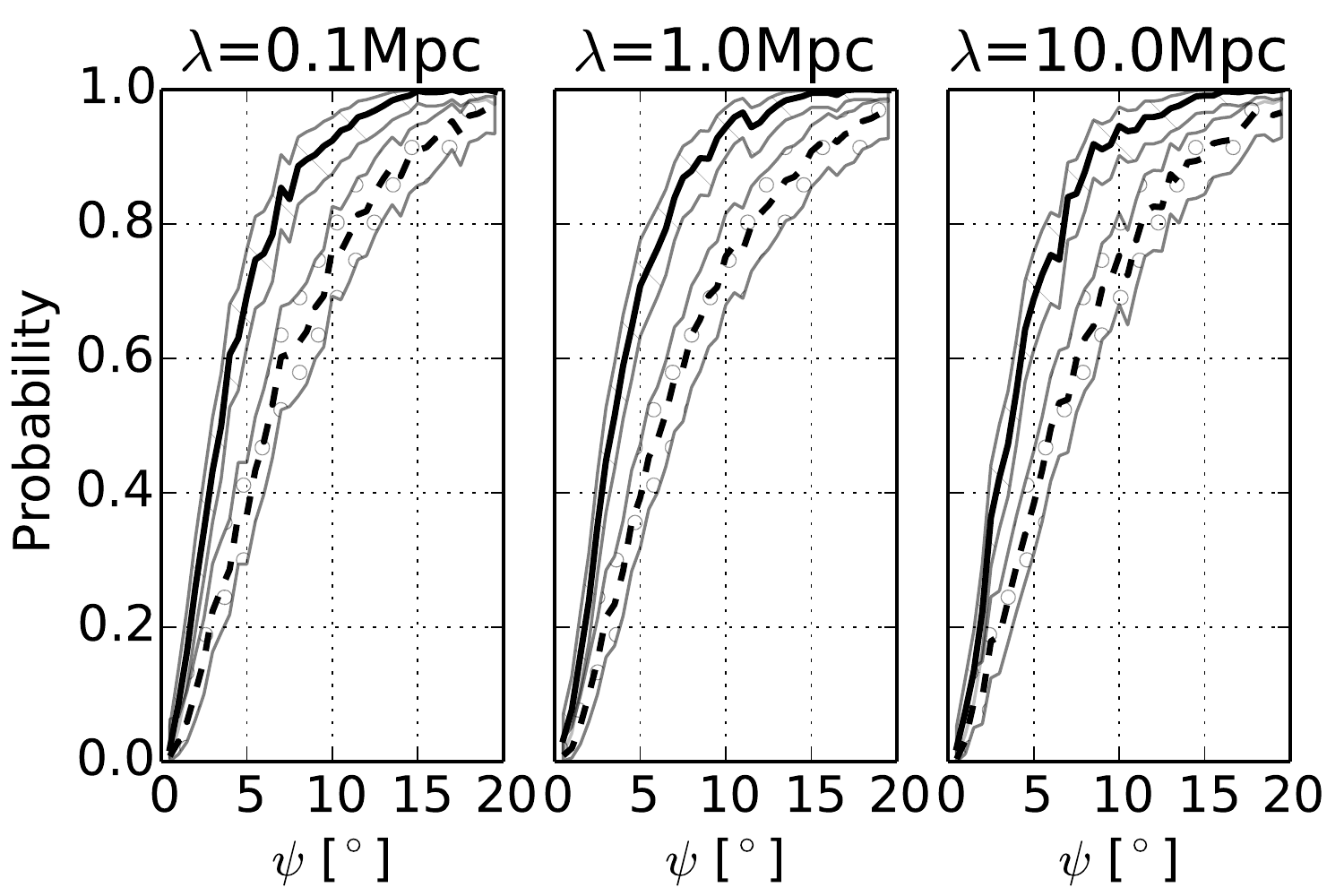}
                \label{dist75}
        
        \caption{Probability of correlation as function of correlation angle $\psi$ for 1000-proton UHECRs, with $0.1$, $1$ and $10$~Mpc IGMF coherence length, scattering in both the galactic halo and disk, and $1^{\circ}$ detector angular resolution. The dashed line is for isotropic events and the solid line is for catalogue sampled events. The shaded regions are $5\sigma$ confidence intervals.}\label{fig:pdata_vs_psi}
\end{figure}

The probability of correlation, $p_{data}$, as a function of correlation angle, $\psi$, for both the VCV catalogue and an isotropic source distribution are shown in Fig.~\ref{fig:pdata_vs_psi}. It is clear that an optimal value of $\psi$ exists that statistically discriminates between source catalogue and isotropic source distribution correlations. In Fig.~\ref{fig:pdata_vs_psi} we have plotted $p_{data}$ vs. $\psi$ assuming different values of the IGMF coherence length $\lambda_B$. We find that there is no significant effect on the optimal value of $\psi$ that depends on $\lambda_B$. 

\clearpage
%For each scenario we consider, we find the maximum probability difference to maximize the possible source correlation significance. Fig.\ref{corre_diffe} illustrates the probability difference between each $1$ Mpc $\lambda_{B}$-line and the assumed isotropic distribution. This plot represents the probability difference relates to Fig.\ref{fig:pdata_vs_psi}. The peak probability shifts to a lower angle for a distance cutoff of 200 Mpc, because when we correlate the scattered events with a higher value of correlation distance the number of AGN sources increases and also their distribution in the sky becomes broader. These two effects bring to a minimization of the correlation angle, as we observe in Fig.\ref{corre_diffe}.  

%The choice of the optimal value of $\psi$ is better represented Fig.~\ref{corre_diffe}. Here we have subtracted the $p_{data}$ for an isotropic source distribution from $p_{data}$ for the catalogue for the $\lambda_{B}$=1~Mpc case. The optimal correlation angle is $\psi=6.3^{\circ}$. 

%Note that in Tables \ref{table:sigmas_1deg} and \ref{table:sigmas_3deg}, the optimal correlation angle, $\psi_{max}$, is below ten degrees in all cases, justifying our use of small scattering angle approximation in Eqns. \ref{eqn:scat1} and \ref{eqn:scat2}. Regardless of the model used for large deflection angles, simulated events with $\psi> 10^{\circ}$ do not correlate to their source. 
 
The optimal value of $\psi$ is found by subtracting the $p_{data}$ for an isotropic source distribution from $p_{data}$ for the catalogue and finding the angle corresponding to the maximum, $\psi_{max}$. Tables  \ref{table:sigmas_1deg} and \ref{table:sigmas_3deg} report the optimal correlation angles for all scenarios considered in this study. The optimal correlation angle is below ten degrees in all cases, justifying our use of small scattering angle approximation in Eqns. \ref{eqn:scat1} and \ref{eqn:scat2}. Regardless of the model used for large deflection angles, simulated events with $\psi> 10^{\circ}$ do not correlate to their source.

\subsection{Results}

\begin{table*}[ht] 
\caption{Statistical significance from isotropy, in units of $\sigma$ for selected proton and iron scenarios, for energy index $\gamma=2.7$, assuming a detector resolution of 1$^{\circ}$. The last column contains the number of sigmas away from isotropic distribution for $100$, $1000$, $10000$ proton UHECR events.} % title of Table 
\centering % used for centering table 
\resizebox{\textwidth}{!}{\begin{tabular}{| c c c  c | c | c c c|} % centered columns (4 columns) 
\hline%\hline
Z &IGMF $\lambda_{B}$[Mpc] & GMF  & Resolution [$^{\circ}$]& $\psi_{max}$[$^\circ$] & N=$10^{2}$ & N=$10^{3}$ & N=$10^{4}$  \\ [0.5ex] % inserts table 
%heading
\hline
1 &	0.1 &	Halo \& Disk  &	1 &	6.36 &	1.1 &	4.0 &	13.5 \\
1 &	0.1 &	Disk Only  &	1 &	4.14 &	2.9 &	11.2 &	39.1 \\
1 &	0.1 &	None  &	1 &	2.78 &	3.9 &	17.4 &	63.1 \\
\hline
1 &	1.0 &	Halo \& Disk  &	1 &	6.41 &	1.2 &	4.1 &	13.3 \\
1 &	1.0 &	Disk Only  &	1 &	4.26 &	2.5 &	10.4 &	36.4 \\
1 &	1.0 &	None  &	1 &	3.35 &	3.3 &	13.9 &	50.0 \\
\hline
1 &	10.0 &	Halo \& Disk &	1 &	6.50 &	1.1 &	3.9 &	13.0 \\
1 &	10.0 &	Disk Only &	1 &	5.03 &	2.1 &	7.7 &	26.1 \\
1 &	10.0 &	None  &	1 &	4.79 &	2.3 &	8.2 &	27.9 \\
\hline
26 &	0.1 &	Halo \& Disk  &	1 &	8.30 &	0.4 &	1.0 &	3.4 \\
26 &	0.1 &	Disk Only  &	1 &	7.90 &	0.6 &	2.0 &	6.6 \\
26 &	0.1 &	None  &	1 &	5.92 &	1.4 &	5.4 &	17.4 \\
\hline
%\hline %inserts single line 
\end{tabular}} 
\label{table:sigmas_1deg} % is used to refer this table in the text 
\end{table*}

\begin{table*}[ht] 
\caption{Statistical significance from isotropy, in units of $\sigma$ for selected proton and iron scenarios, for energy index $\gamma=2.7$, assuming a detector resolution of 3$^{\circ}$. The last column contains the number of sigmas away from isotropic distribution for $100$, $1000$, $10000$ proton UHECR events.} % title of Table 
\centering % used for centering table 
\resizebox{\textwidth}{!}{\begin{tabular}{| c c c c  | c | c c c|} % centered columns (4 columns) 
\hline%\hline
Z &IGMF $\lambda_{B}$[Mpc] & GMF & Resolution [$^{\circ}$]& $\psi_{max}$[$^\circ$] & N=$10^{2}$ & N=$10^{3}$ & N=$10^{4}$  \\ [0.5ex] % inserts table 
%heading
\hline
1 &	0.1 &	Halo \& Disk &	3 &	6.51 &	1.2 &	4.0 &	13.1 \\
1 &	0.1 &	Disk Only &	3 &	4.95 &	2.2 &	8.4 &	28.4 \\
1 &	0.1 &	None  &	3 &	4.64 &	2.6 &	9.2 &	31.5 \\
\hline
1 &	1.0 &	Halo \& Disk &	3 &	6.53 &	1.2 &	3.8 &	13.1 \\
1 &	1.0 &	Disk Only &	3 &	5.04 &	1.9 &	8.0 &	27.1 \\
1 &	1.0 &	None  &	3 &	4.75 &	2.2 &	8.8 &	30.1 \\
\hline
1 &	10.0 &	Halo \& Disk &	3 &	6.58 &	1.0 &	3.9 &	12.9 \\
1 &	10.0 &	Disk Only  &	3 &	5.48 &	1.8 &	6.7 &	22.4 \\
1 &	10.0 &	None  &	3 &	5.36 &	2.0 &	6.8 &	23.6 \\
\hline
26 &	0.1 &	Halo \& Disk  &	3 &	8.30 &	0.3 &	0.9 &	3.3 \\
26 &	0.1 &	Disk Only  &	3 &	7.90 &	0.6 &	2.0 &	6.5 \\
26 &	0.1 &	None  &	3 &	6.07 &	1.4 &	4.8 &	16.4 \\
\hline
\end{tabular}} 
\label{table:sigmas_3deg} % is used to refer this table in the text 
\end{table*} 

Tables \ref{table:sigmas_1deg} and \ref{table:sigmas_3deg} lists the significance of the correlation to the parent catalogue over an isotropic distribution for several possible IGMF coherence lengths, galactic magnetic field models, and composition models. The significances are computed assuming a $1^{\circ}$ detector angular resolution in Table \ref{table:sigmas_1deg} and a $3^{\circ}$ detector angular resolution in Table \ref{table:sigmas_3deg}.
 
There are several proton-dominated scenarios which result in highly significant source catalogue correlations. Inclusion of galactic magnetic field scattering reduces the likelihood of a significant detection ($>5\sigma$) the most. On the order of 1,000 events are required for scenarios with only protons and full scattering off the galactic halo and disk to discriminate between anisotropic and isotropic distribution of UHECRs at the $5\sigma$ level. In all proton cases considered, scattering in the halo substantially reduces the significance of a detection of correlations.

There are no realistic scenarios wherein iron UHECRs result in significant correlations with the catalogue. We only report the results for $\lambda_{B}=0.1$~Mpc, because the results for longer coherence lengths are insignificant except for 10,000 UHECRs that experience no scattering in the galactic disk or halo. Even the optimistic case that assumes no scattering within the galaxy, small (0.1~Mpc) IGMF coherence lengths, and a sample size of 1000 events shows correlations only at the $4-5\sigma$ level. If scattering in the disk is included, then at least several thousand events are required to detect significant correlations. If scattering in the halo is also included, then no significant departure from isotropy is found for any number of events considered. %Studies of the scattering off of the galactic magnetic field are relatively recent \cite{neronov_2009,jansson_2009}, and we confirm their results that iron nuclei are likely to be isotropized by the galactic magnetic fields. 

The effects of varying the coherence length are most pronounced in scenarios where we do not include galactic magnetic field scattering. In such cases, it is clear that longer coherence lengths scatter both protons and iron more, making them more consistent with isotropy. For cases that include scattering in the galactic disk, the correlation significance of 1000 detected protons can be up to 4$\sigma$ higher for $\lambda_{B}$ of 0.1 Mpc than for 10 Mpc. However, when including scattering in both the disk and the halo, galactic magnetic field scattering dominates over the IGMF contribution, such that the correlation significances vary by less than 1$\sigma$ for all coherence lengths. 

%Our results did not change significantly when using a detector with angular resolution of $3^{\circ}$. 
We find that an improved detector resolution does not translate directly to improved sensitivity to detecting a source catalogue correlation at greater than 5$\sigma$. In the simplest cases without galactic magnetic field scattering, $\psi_{max}$ is reduced when the angular resolution is improved, thereby increasing the correlation significance. With scattering in the galactic disk, the improved angular resolution can reduce the required number of events to reach the $5\sigma$ level by a factor of a few. This effect degrades with increasing coherence length. However, when scattering in the halo is included, the differences in correlation significance are consistent with statistical fluctuations at the $<1 \sigma$ level.

\section{Conclusions}
The streamlined model presented here places a lower bound on the requirements for a state-of-art experiment to detect significant correlations from a source catalogue. We analyzed the required event rate above an energy threshold of 60~EeV of a future UHECR all-sky instrument for identifying sources, assuming several realistic scenarios with differing cosmic ray composition and magnetic field models. Such a simple parametric simulation does not require large computing power, but is capable of characterizing the trends and challenges for source identification. We have assumed perfect energy resolution, which neglects the effect of cosmic rays below our energy threshold spilling over into the data sample, also making our lower bound optimistic. 

%Our focus is to characterize the sensitivity of an experiment to several assumed model parameters, especially the relatively unconstrained intergalactic and galactic magnetic fields. 

We find that when both the halo and disk magnetic fields are included in our scattering model, angular resolutions better than $3^{\circ}$ do not greatly improve the detectability of the source catalogue correlations. The optimal correlation angle is greater than a few degrees in cases that include galactic magnetic field scattering. This is consistent with Eqns. \ref{eqn:scat1} and \ref{eqn:scat2} that scatter events with energies less than 100 EeV by several degrees. Combined with the power-law flux of UHECRs, most events will arrive at Earth scattered by more than $3^{\circ}$. Therefore, our results indicate that an improved angular resolution of $1^{\circ}$ is not expected to significantly improve source catalogue correlations with our current understanding of scattering in the galactic halo. %The peak correlation angle varies with magnetic field model, but it is $\ge2^{\circ}$ in all cases. Even when neglecting scattering in the galaxy, the $\psi$ resulting in maximum correlation indicates that a detector's minimum angular resolution need only be as good as $\sim2^{\circ}$, even in the case of short $\lambda_{B}$.% Long coherence lengths have better peak correlation angles, but the reduce the anisotropy in the arrival directions of cosmic rays as they increase the scattering angle.

Scattering in the galactic disk isotropizes the cosmic-ray distribution more than scattering outside of the galaxy, despite the relatively unconstrained intergalactic magnetic field coherence length. Longer coherence lengths of the IGMF scatter UHECRs even further, but that scattering does not dominate the results of the correlation analysis. Future experiments would benefit from an improved understanding of the magnetic fields within the galaxy. This is underlined by how strongly the inclusion of scattering in the galactic halo affects the significance of a correlation.

If cosmic rays are predominantly iron, a detection of greater than $10^{4}$ events above 60~EeV  would be required for source identification, and therefore, an exposure greater than 100 times the state of the art. However, if they are predominately proton, an experiment that detects $10^{3}$ could expect correlations with a source catalogue at $>4\sigma$ even with deflection in the galactic halo and disk. This implies that a full sky survey of UHECRs should have an exposure at least 10 times the current state-of-the-art as well as an improved understanding of the composition of cosmic rays, which is consistent with the conclusions of \cite{Oikonomou_2015}. % at those energies are required to distinguish among astrophysical models. 
Given the negative redshift distribution of the VCV catalogue, as used in this study, this is likely an optimistic lower bound.
The simulations presented here indicate that the lack of a significant correlation of cosmic rays with energy $>60$~EeV to nearby VCV catalogue sources in both the Auger and TA experiments does not exclude the possibility that AGN are the acceleration sites of UHECRs.
%Outlook
%\begin{itemize}
%\item {It would be interesting to look at the significance of correlation as a function of distance cutoff as a statistical means of discriminating between protons and iron nuclei}
%\item {It would be interesting to perform this analysis with different catalogues (GRB, x-ray bright AGN, pulsars). How generic is VCV?}
%\end{itemize}

\appendix
\section{Number of Cosmic Rays Arriving from a Catalogue}
We calculate the number of cosmic rays observed at Earth from sources in a catalogue by assuming a specific intensity at the source, total exposure on a given source, and tracking the energy losses due to propagation and cosmological expansion. We start from the specific intensity $I_{i}$, given as the number per unit time, area, solid angle, and energy, also known as differential flux, of cosmic rays arriving from source $i$ to the observer. This is defined according to the differential relation
\begin{equation}
dN=
I_{i,obs} 
dt_{obs}dE_{obs}\cos\theta_{i,obs}d\sigma_{obs}d\Omega_{i,obs}
\end{equation}
The $dt_{obs}$ term gives the arrival rate of UHECRs of energy between $E_{obs}$ and $E_{obs}+dE_{obs}$ in a detector differential area element $d\sigma_{obs}$ pointed at angle $\theta_{obs}$ with respect to the source which subtends a differential solid angle $d\Omega_{i,obs}$ in the sky. The specific intensity is, in general, a function of the surface of the detector $\mathbf{r}_{obs}$, and direction of observation $\theta_{obs}$, $\phi_{obs}$ with respect to the source position.

The relation between the specific intensity of a source as seen by the observer and that as seen from the source frame of reference is given by
\begin{equation}
I_{i,obs} 
= 
\frac{dt_{src}}{dt_{obs}} \
\frac{dE_{src}}{dE_{obs}} \
\frac{\cos\theta_{src}d\sigma_{src}}{\cos\theta_{obs}d\sigma_{obs}} \
\frac{d\Omega_{src}}{d\Omega_{obs}} \
I_{i,src} 
\end{equation}
with the specific intensity of the source illuminating the detector $I_{i,src}$ defined by the relation
\begin{equation}
dN=
I_{i,src} 
dt_{src}dE_{src}\cos\theta_{i,src}d\sigma_{src}d\Omega_{i,src}
\end{equation}

The number of cosmic rays arriving at a detector with energy $E_{obs}$ greater than $E_{cut}$ from a catalogue with $M$ sources is given by
\begin{equation}
N(E_{obs}>E_{cut}) = 
\sum_{i=1}^{M} 
\int_{0}^{T_{i}}dt_{obs} 
\int_{0}^{\infty}dE_{obs}\Theta(E_{obs}-E_{cut})
\int_{A_{det}}
\int_{\Omega_{i,obs}}
\cos\theta_{i,obs}d\sigma_{obs}d\Omega_{i,obs}
\ I_{i,obs}
\label{eqn:num_obs}
\end{equation}

The source time and observer time are related by $dt_{src}/dt_{obs}=(1+z_i)$, which gives the time dilation of the source emission rate. The second term $dE_{src}/dE_{obs}$ accounts for the change in spectral band due to energy propagation effects. The \'{e}tendue of the source $dG_{src}=\cos\theta_{src}d\sigma_{src}d\Omega_{src}$ is the differential emission surface area of the source with a cosine projection factor on the solid angle subtended by the detector, in the frame of reference of the source. This is related to the \'{e}tendue of the observer $dG_{obs}=\cos\theta_{obs}d\sigma_{obs}d\Omega_{src}$ via Etherington's~\cite{etherington_1933} reciprocity theorem~$dG_{obs} = (1+z)^2 dG_{src}$. We note that the distance between the source and the observer from the point of view of the source is the comoving distance, while from the point of view of the observer, it is the angular diameter distance.

We may rewrite Equation~\ref{eqn:num_obs} as
\begin{equation}
N(E_{obs}>E_{cut}) = 
\sum_{i=1}^{M} 
\frac{T_{i}}{1+z_i} 
\int_{0}^{\infty}dE_{src}\Theta(E_{obs}(E_{src})-E_{cut})
\int_{A_{src}}
\int_{\Omega_{i,src}}
\cos\theta_{i,src}d\sigma_{src}d\Omega_{i,src}
\ I_{i,src}
\end{equation}
where we have integrated over the observer time $t_{obs}$ to arrive at a total exposure time $T_k$. We have changed variables to $E_{obs}$, which is a function of $E_{src}$. The \'{e}tendue is integrated over $A_{src}$, which is the area of emission of the source and over the solid angle subtended by the detector as seen by the source $\Omega_{i,src}$. 

We assume the specific intensity of each source to follow the same model given by
\begin{equation}
I_{i,src} = 
I_0 \ f(\mathbf{r}_{src}, \theta_{src}, \phi_{src}) \ \left(\frac{E_{src}}{E_{0}}\right)^{-\gamma_g}.
\end{equation}
The scalar $I_0$ sets the level of intensity. The function $f(\mathbf{r}_{src}, \theta_{src}, \phi_{src})$ sets the emission beam pattern of the source, which may, in general, depend on the location of the surface of the source $\mathbf{r}_{src}$ and direction of emission $\theta_{src},\phi_{src}$. We may write the specific intensity integrated over the source \'{e}tendue as 
\begin{equation}
\int_{A_{src}}
\int_{\Omega_{i,src}}
\cos\theta_{i,src}d\sigma_{src}d\Omega_{i,src}
\ I_{i,src}
=
I_0 \left(\frac{E_{src}}{E_{0}}\right)^{-\gamma_g}
\int_{A_{src}}
\int_{\Omega_{i,src}}
\cos\theta_{i,src}d\sigma_{src}d\Omega_{i,src}
\ f(\mathbf{r}_{src}, \theta_{src}, \phi_{src})
\end{equation}
to give
\begin{equation}
\int_{A_{src}}
\int_{\Omega_{i,src}}
\cos\theta_{i,src}d\sigma_{src}d\Omega_{i,src}
\ I_{i,src}
=
I_0 A_{eff,src} \frac{A_{eff,det}}{4\pi d_C^2(z_i)}  
\left(\frac{E_{src}}{E_{0}}\right)^{-\gamma_g}
\end{equation}

The solid angle subtended by the detector from the view of the source is $\Omega_{i,src}=A_{eff,det}/\left(4\pi d_C^2(z_i)\right)$. The effective area of the source $A_{eff,src}$ is the emission area referenced to an isotropic emitter (as is typically done in antenna theory). We write $L_0=I_0A_{eff,src}$ and assume the effective area of the source is the same for each source in the catalogue. We write the source luminosity as
\begin{equation}
L_{src}(E_{src})
=
L_0 \left(\frac{E_{src}}{E_{0}}\right)^{-\gamma_g}
\end{equation}
to give
\begin{equation}
N(E_{obs}>E_{cut}) = 
\sum_{i=1}^{M} 
\frac{T_{i}}{1+z_i} 
\frac{A_{eff,det}}{4\pi d_C^2(z_i)}
\int_{0}^{\infty}dE_{src}\Theta(E_{obs}(E_{src})-E_{cut})L(E_{src})
\end{equation}
Finally, we assume the exposure to each source is the same value given by $TA_{eff,det}$ and factor it out of the sum over all the catalogue sources to give Equation~\ref{eqn:num_events}.

\section{Energy Loss Propagator}
The energy loss propagator derivation follows the approach presented in~\cite{berezinsky_2006}. The calculation has been adapted to make use of the energy loss length curves presented in Section~\ref{sec:loss_length}. The energy losses associated with propagation has contributions from cosmological redshift as well as interaction with the photon background. Given that the photon background density also changes with redshift we model the energy change with redshift $z$ as 	
\begin{equation}
\frac{dE}{dz} = \left(\frac{dE}{dz}\right)_{redshift}+\left(\frac{dE}{dz}\right)_{\gamma}
\label{eqn:eloss}
\end{equation}
where the energy losses due to redshift are given by
\begin{equation} 
\left(\frac{dE}{dz}\right)_{redshift}=(1+z)^{-1}E
\end{equation}
The losses due to photon background interactions ($\gamma$) are given by
\begin{equation} 
\left(\frac{dE}{dz}\right)_{\gamma}=\left(\frac{dE}{dx}\right)_{\gamma}\frac{dx}{dz}
\end{equation}
where the energy loss is
\begin{equation} 
\left(\frac{dE}{dx}\right)_{\gamma}=-\frac{1}{\lambda_{\gamma}}E,
\end{equation}
where $\lambda_{\gamma}$ is the loss length described in Section~\ref{sec:loss_length}. The differential change in comoving distance $dx/dz$ is given by 
\begin{equation} 
\frac{dx}{dz}=-\frac{c}{H(z)}.
\end{equation}
The Hubble parameter is
\begin{equation} 
H(z)=H_{0}\sqrt{\Omega_M(1+z)^3+\Omega_{\Lambda}}
\end{equation}
for a $\Lambda$CDM cosmology with $H_0$=72~km/Mpc/s, $\Omega_M=0.3$, and $\Omega_{\Lambda}=0.7$.

The loss length $\lambda_{\gamma}$ also has cosmological corrections. The loss length $\lambda_{\gamma}(E)$ for a cosmic ray with energy $E$ at $z=0$ is given by
\begin{equation} 
\frac{1}{\lambda_{\gamma}(E)}=\int_{0}^{\infty}dE_{\gamma}n_{\gamma}(E_{\gamma})\frac{d\sigma}{dE_{\gamma}}(E, E_{\gamma})
\end{equation}
Where $\frac{d\sigma}{dE_{\gamma}}(E, E_{\gamma})$ is the differential interaction cross-section of a cosmic ray of energy $E$ with a photon of energy $E_{\gamma}$ and $n_{\gamma}(E)$ is the photon density as $z=0$. The loss length $\lambda'_{\gamma}(E)$ for a cosmic ray with energy $E$ at a different redshift $z$ is given by
\begin{equation} 
\frac{1}{\lambda'_{\gamma}(E)}=\int_{0}^{\infty}dE'_{\gamma}n'_{\gamma}(E'_{\gamma})\frac{d\sigma}{dE'_{\gamma}}(E, E'_{\gamma}).
\end{equation}
The photon density at redshift $z$ is given by $n'(E'_{\gamma})=(H(z)/H_0)^3n(E_{\gamma})$. Substitution of variables of $E'_{\gamma}$ for $E_{\gamma}$ in the integrand gives
\begin{equation}
\lambda'_{\gamma}(E) = \left(\frac{H_{0}}{H(z)}\right)^3\lambda_{\gamma}(E).
\end{equation}

Combining the terms above into Equation~\ref{eqn:eloss} we arrive at
\begin{equation}
\frac{dE}{dz}=\frac{1}{1+z}E + \frac{1}{\lambda_{\gamma}(E)}\frac{cH_{0}^{3}}{H^2(z)}E
\end{equation}

We can discretize this equation for the energy $E_{j}$ and redshift $z_j$ at step $j$, the energy $E_{j+1}$ at step $j+1$ with redshift $z_{j+1} = z_{j} - \Delta z$ is given by 
\begin{equation}
E_{j+1}=E_{j}\left(1 - \frac{\Delta z}{1+z_j} - \frac{\Delta z}{\lambda_{\gamma}(E_j)}\frac{ cH^3_0}{H^2(z_j)}  \right)
%-\frac{\Delta z H^3_0}{H^2(z_k)}\frac{c}{\lambda_{\gamma}(E_k)}
\end{equation}

\acknowledgments
This research was carried out, in part, at the Jet Propulsion Laboratory,
California Institute of Technology, under a contract with the National
Aeronautics and Space Administration. Copyright 2014. All rights
reserved.

%\section*{Bibliography}
\bibliographystyle{elsarticle-num}
\bibliography{<your-bib-database>}

%% Authors are advised to submit their bibtex database files. They are
%% requested to list a bibtex style file in the manuscript if they do
%% not want to use elsarticle-num.bst.

%% References without bibTeX database:
%%%%%%%%%%%%%%%%%%%%%%%%%%%%%%%%%%%%%%%%%%%%%%%%%%%%%%%%%%%%%%%%%%%%%%%% 
%%%%%%%%%%%%%%%%%%%%%%%%%%%%%%%%%%%%%%%%%%%%%%%%%%%%%%%%%%%%%%%%%%%%%%%% 
%%%%%%%%%%%%%%%%%%%%%%%%%%%%%%%%%%%%%%%%%%%%%%%%%%%%%%%%%%%%%%%%%%%%%%%% 

% format from an astroparticle article:
% Author, et al., Journal Volume (Year) Page Number.
%  D.J. Bird, et al., Phys. Rev. Lett. 71 (1993) 3401.
% Authors, Volume (Year) p. 167
% J.A. Bellido, R.W. Clay, B.R. Dawson, M. Johnston-Hollitt Astropart. Phys., 15 (2001), p. 167

\end{document}